\documentclass[twocolumn,showpacs,preprintnumbers,amsmath,amssymb]{revtex4}
\usepackage{graphicx}% Include figure files
\usepackage{dcolumn}% Align table columns on decimal point
\usepackage{bm}% bold math

\def\bmatrix{\left[\begin{array}}
\def\ematrix{\end{array}\right]}
\def\massll{(\widetilde{M}_d^2)_{LL}}
\def\massrr{(\widetilde{M}_d^2)_{RR}}

\def\vubus{V_{ub}V_{us}^\ast}

\def\vtbts{V_{tb}V_{ts}^\ast}

\def\cvcbcs{V_{cb}^\ast V_{cs}}
\def\cvtbts{V_{tb}^\ast V_{ts}}

\begin{document}

%\preprint{APS/123-QED}

\title{\boldmath
Effect of Supersymmetric Right-handed Flavor Mixing
on $B_s$ decays }
\vfill
\author{Wei-Shu Hou}
\author{Makiko Nagashima}%
%\affiliation{%
%\author{Makiko Nagashima}
%\affiliation{${}^a$Institute of Physics, Academia Sinica,
%                 Taipei, Taiwan 115, R.O.C.}
\affiliation{Department of Physics, National Taiwan
University, Taipei, Taiwan 106, R.O.C. }%
%

%\date{\today}
%
%\vskip -1cm
%
\vfill
\begin{abstract}
Motivated by the possibility of ${\cal S}_{\phi K_S}<0$, we study
the implications for $B_s$ meson system. In a specific model that
realizes ${\cal S}_{\phi K_S}<0$ with large $s$-$b$ mixing,
right-handed dynamics and a new CP phase, we present predictions
for CP asymmetries in $B_s\to J/\psi\phi$, $K^+K^-$ and
$\phi\gamma$ decays. Even if the measurement of time-dependent CP
asymmetry becomes hampered by very fast $B_s$ oscillation, a
finite difference between the decay rates of $B_s$ mass
eigenstates may enable the studies of CP violations with untagged
data samples. Thus, studies of CP violation in the $B_s$ system
would remain useful for the extraction of new physics information.
\end{abstract}
\pacs{11.30.Er, 11.30.Hv, 12.60.Jv, 13.25.Hw}
\maketitle

%\protect \textbackslash\textbackslash}

\section{Introduction}
%%%
The current experimental average of mixing-dependent CP asymmetry
in $B_d\to \phi K_S$ decay is ${\cal S}_{\phi K_S}= -0.15\pm 0.33$
\cite{Browder}. This constitutes a $2.7\sigma$ deviation from
$\sin 2\phi_1 = 0.736\pm 0.049$ as measured in $B_d\to J/\psi K_S$
and other charmonium channels \cite{Browder}. The standard model
(SM) asserts ${\cal S}_{\phi K_S} = \sin 2\phi_1$ to the per cent
level. Although there is disagreement between the measurements of
Belle \cite{Belle03} and ${\rm B_{A}B_{AR}}$ \cite{Browder} at
$2.1\sigma$ level at present, the new physics hint may well be
real. On the other hand, the SM also asserts ${\cal
S}_{K_S\pi^0}$, ${\cal S}_{\eta^\prime K_S} = \sin 2\phi_1$, which
seem to be supported by the experimental values. Many new physics
(NP) scenarios have been advanced. In particular, it has been
pointed out that the new physics interaction should be
right-handed so that, while ${\cal S}_{\phi K_S}$ can be negative
in sign, ${\cal S}_{\eta^\prime K_S}$ \cite{Kagan,KK} and ${\cal
S}_{K_S\pi^0}$ \cite{CHNphik} $\sim \sin 2\phi_1\;>0$ can be
maintained.

%%%
In our previous work \cite{CHNphik}, we have studied the
implications of ${\cal S}_{\phi K_S} < 0$ for charmless $B_d$
decay modes, in the framework of Abelian flavor symmetry (AFS)
combined with supersymmetry (SUSY). We assume that SUSY is at TeV
scale, and the AFS scale is not far above the SUSY scale. For sake
of clarity, we introduced three parameters as follows: gluino mass
$m_{\tilde g}$, light right-handed squark $\widetilde{sb}_1$ mass
$\tilde m_1$ (not independent from squark mass scale $\tilde m$)
and just one extra CP violating phase $\sigma$. Taking into
account contributions from the chromo-dipole operator, we found
that large $\tilde s_R$-$\tilde b_R$ mixing can turn ${\cal
S}_{\phi K_S}$ negative for $\sigma\sim 40^\circ$-$90^\circ$,
while for $\sigma\sim 180^\circ$-$360^\circ$, ${\cal S}_{\phi
K_S}$ is larger than the SM value of $0.74$. Combining $b\to
s\gamma$ and $B\to \phi K_S$ rates and ${\cal S}_{\phi K_S}<0$,
the current experimental results seem to suggest $\sigma\sim
65^\circ$. We now extend our study to the $B_s$ system.

%%%
The present bound of the mass difference $\Delta m_{B_s}$ is
around 15 ${\rm ps}^{-1}$. Compared to $\Delta m_{B_d}=0.502$
${\rm ps}^{-1}$ \cite{PDG}, $B^0_s$-$\bar B^0_s$ oscillations are
already very rapid. As pointed out in \cite{CHNphik}, the large
effect of $S_{\phi K_S}<0$ calls for rather light
$\widetilde{sb}_1$, while $m_{\tilde g}$ cannot be too heavy. It
is then found that $\Delta m_{B_s}\gtrsim 70$ ${\rm ps}^{-1}$ is
hard to avoid. Measurement of such fast oscillations at Tevatron
Run II now appears hopeless, and it would be challenging even for
the next generation LHCb and BTeV experiments. While finding
$\Delta m_{B_s} > \Delta m_{B_s}^{\rm SM}$ itself would constitute
evidence beyond SM, a better quantity for revealing NP is the weak
phase of $B^0_s$-$\bar B^0_s$ mixing, {\it i.e.} $\sin
2\Phi_{B_s}$. Since the SM predicts $\sin 2\Phi_{B_s}\approx 0$
due to the absence of CKM phase in $\cvtbts/\vtbts$, the
measurement of $\sin 2\Phi_{B_s}$ is expected to be a good NP
probe. We found that, for $\sigma\sim 65^\circ$, $\sin
2\Phi_{B_s}$ can vary over a very wide range \cite{CHNphik}.
Although the observation of $\sin 2\Phi_{B_s}$ itself may be
hampered by the very fast $B_s$ oscillation, measurement of $\sin
2\Phi_{B_s}$ would shed further light on $\widetilde{sb}_1$
parameters.

%%%
Whether the measurement of mixing-dependent CP asymmetries in the
$B_s$ system becomes very challenging or not, it is useful to
remember that untagged data may provide an alternative handle for
studies of CP violation in the $B_s$ system.
The general formula for time-dependent CP asymmetry in $B_{q}\to
f$ decay, where $f$ is a CP eigenstate, is given by
\cite{Hagelin,Dighe,Fleischer,TevatronII}
\begin{eqnarray}
& & a_{\rm CP}(B_{q}(t)\to f)=
\frac{\Gamma (\bar B_{q}(t)\to f)-\Gamma (B_{q}(t)\to f)}
{\Gamma (\bar B_{q}(t)\to f)+\Gamma (B_{q}(t)\to f)}  \nonumber \\
\nonumber \\
&&=
 \frac{{\cal A}_f\cos \Delta m_{B_q}t + {\cal S}_f\sin
\Delta m_{B_q}t} {\cosh\frac{\Delta\Gamma_q t}{2} +{\cal
A}_{\Delta\Gamma}\;\sinh\frac{\Delta\Gamma_q t}{2}
},
\label{eq:timeCPI}
\end{eqnarray}
where ${\cal A}_f$, ${\cal S}_f$ and
${\cal A}_{\Delta\Gamma}$ are expressed in terms of
decay amplitudes $A(B\to f)$ and $\bar A(\bar B\to f)$ as
\begin{eqnarray}
& &{\cal A}_f=\frac{|\bar A|^2-|A|^2}
{|\bar A|^2+|A|^2},
\hspace{5mm}
{\cal S}_f=
\frac{2\xi{\rm Im}\bigl[\frac{q}{p}\bar A A^\ast\bigr]}
{|\bar A|^2+|A|^2},
\nonumber \\
& &{\cal A}_{\Delta\Gamma}= \frac{2\xi{\rm
Re}\bigl[\frac{q}{p}\bar A A^\ast\bigr]} {|\bar A|^2+|A|^2},
\label{eq:timeCPII}
\end{eqnarray}
and $\vert {\cal A}_f\vert^2 + \vert {\cal S}_f\vert^2 + \vert {\cal
A}_{\Delta\Gamma} \vert^2 = 1$. The final state $f$ satisfies
${\cal CP}|f\rangle =\xi |f\rangle$, and $q/p$ is related to the
weak phases describing $B^0_q$-$\bar B^0_q$ mixing. Throughout
this paper, we set the mass difference $\Delta m_{B_q}>0$ and
define the width difference $\Delta\Gamma_q$ as follows,
\begin{eqnarray}
& & \Delta m_{B_q}=m_{B_{qH}}-m_{B_{qL}}, \nonumber \\
& & \Delta\Gamma_q=\Gamma_{B_{qH}}- \Gamma_{B_{qL}},
\end{eqnarray}
where $B_{qH}^0$ and $B_{qL}^0$ denote the mass eigenstates of the
$B^0_q$ system, and  $\Gamma_q$ is the average width. The mass
eigenstates can be approximated by the CP eigenstates. However,
for the $B_s$ system the correspondence is still blind.

%%%
For the $B_d$ system, one has $\Delta\Gamma_{d}/\Gamma_d \to 0$,
and Eq.~(\ref{eq:timeCPI}) simplifies to the more familiar form of
\begin{eqnarray}
a_{\rm CP}(B_{d}(t)\to f)=
{\cal A}_f\cos \Delta m_{B_d}t + {\cal S}_f\sin \Delta m_{B_d}t,
\end{eqnarray}
where ${\cal A}_f$ and ${\cal S}_f$ can be determined by measuring time
distributions of flavor tagged data. However, very rapid
$B^0_s$-$\bar B^0_s$ oscillations, {\it i.e.} $\Delta m_{B_s}\gg
\Gamma_s$, may prevent us from distinguishing between an initial
$B_s^0$ or $\bar B_s^0$.
Fortunately, the quantity ${\cal A}_{\Delta\Gamma}$ also probes CP
violation (as can be seen from Eq.~(\ref{eq:timeCPII})), and can
be extracted by using {\it untagged} $B_q$ data samples,
\begin{eqnarray}
\Gamma[f,t] &\equiv& \Gamma (B_q(t)\to f)+\Gamma (\bar B_q(t)\to
f) \nonumber \\
& \propto&  e^{-\Gamma_qt}\;(|A|^2+|\bar A|^2)
\cosh\frac{\Delta\Gamma_q t}{2} \nonumber \\
& & \times
\biggl\{1+{\cal A}_{\Delta\Gamma}\tanh\frac{\Delta\Gamma_q t}{2}\biggr\}.
\label{eq:untagg}
\end{eqnarray}
Unlike $\Delta\Gamma_{d}/\Gamma_d \to 0$ for $B_d$,
$\Delta\Gamma_{s}/\Gamma_s$ may be as large as $O(10\%)$
\cite{Beneke}. This makes the study of CP violation in $B_s$
system hopeful, even if very rapid $B^0_s$-$\bar B^0_s$
oscillations prevents us from extracting ${\cal S}_f$ from flavor
tagged data.

%%%
All quantities in Eq.~(\ref{eq:untagg}) can in principle be
measured at hadron colliders. The average width,
$\Gamma_s=(\Gamma_{B_{s}}^{(+)}+\Gamma_{B_{s}}^{(-)})/2$, can be
measured via flavor-specific studies such as semileptonic decays,
where both CP even and odd widths $\Gamma_{B_{s}}^{(\pm)}$ are
present. $\Gamma_{B_{s}}^{(+)}$ can be measured via decays such as
$B_s\to J/\psi\phi$, which is dominantly CP-even \cite{CDF}. Thus,
one can in principle obtain
$|\Delta\Gamma_s/2|=|\Gamma_s-\Gamma_{B_{s}}^{(+)}|$. It is
expected that studies of $B_s$ decays can determine
$\Delta\Gamma_s/\Gamma_s$ up to a few \% accuracy
\cite{TevatronII}. If $\Delta\Gamma_s/\Gamma_s$ is around $10\%$
as asserted, studies with untagged data can be a promising way to
measure CP violation.

%%%
As $B_s$ studies can help confirm and shed further light on new
physics that might already be emerging from $B_d$ studies, in this
paper we study the impact of light $\widetilde{sb}_1$ on $B_s$
meson decays, with special attention to a new CP phase $\sigma\sim
65^\circ$. We study the measurements that can assist in
determining the model parameters. The outline of this paper is as
follows. In Section~\ref{sec:model}, we recall the necessary
features of the new physics model with maximal mixing between
right-handed squarks $\tilde s_R$ and $\tilde b_R$. Results of our
analysis are shown in Section~\ref{sec:result}, for $B_s\to
J/\psi\phi$, $K^+K^-$ and $\phi\gamma$. Conclusion is given in
Section~\ref{sec:discussion}.

\section{\label{sec:model}
maximal $\tilde s_R$-$\tilde b_R$ squark mixing}
Within SM we have no direct knowledge of right-handed quark
mixings, since the weak interaction probes just the left-handed
sector. On the other hand, we are still far from a solution to the
flavor problem. SUSY can in principle bring forth extra
right-handed dynamics, but it does not address the flavor issue.
As one tries to understand the right-handed flavor sector by
assuming AFS, near maximal $s_R$-$b_R$ mixing can be realized.
With supersymmetric AFS, furthermore, maximal $\tilde s_R$-$\tilde
b_R$ squark mixing \cite{CH,ACH01} brings forth an extra CP
violating phase and $s_R\tilde b_R\tilde g$ couplings. As pointed
out in \cite{ACH01}, decoupling the $d$ flavor is preferred to
evade the constraints from kaon system as much as possible.

With $d$ flavor decoupled, we can reduce $3\times 3$ down quark
mass matrix to $2\times 2$ submatrix relevant to $s$ and $b$
flavors. Normalized to $m_b$, the down quark mass matrix element
$M^{(d)}_{ij}$ has the diagonal terms $M^{(d)}_{33}\simeq 1$ and
$M^{(d)}_{22}\simeq \lambda^2$, where $\lambda\simeq 0.22$ is the
Wolfenstein parameter. Taking analogy with $V_{cb} \simeq
\lambda^2$ gives $M^{(d)}_{23} \simeq \lambda^2$, but
$M^{(d)}_{32}$ is unknown for lack of right-handed flavor
dynamics. With effective AFS \cite{Nir}, the {\it Abelian} nature
implies $M^{(d)}_{23}M^{(d)}_{32}\sim M^{(d)}_{22}M^{(d)}_{33}$,
hence $M^{(d)}_{32}\sim 1$ is deduced. This may be the largest
off-diagonal term, but its effects is hidden from our view within
the SM. However, introducing SUSY, its effect may be revealed via
the super-partners of right-handed quarks.

Decoupling $d$ flavor reduces the down squark mass matrix from
$6\times 6$ to $4\times 4$. Focusing on the $2\times 2$ $RR$
submatrix $\massrr^{(sb)}$, by the hermitian nature one finds just
one extra CP violating phase, denoted as $\sigma$, and we
parameterize $\massrr^{(sb)}$ as,
\begin{eqnarray}
\massrr^{(sb)} &=& \bmatrix{cc}\widetilde m^2_{22}&
\widetilde m^2_{23}e^{-i\sigma} \\
\widetilde m^2_{23}e^{i\sigma} & \widetilde m^2_{33} \ematrix\;.
\label{eq:sbbase}
\end{eqnarray}
The squark mass eigenvalues are $\tilde m^2_{1,2}= (\tilde
m^2_{22}+\tilde m^2_{33} \mp \sqrt{(\tilde m^2_{22}-\tilde
m^2_{33})^2 +4\tilde m^4_{23}})/2$, which are reached by
diagonalizing $\massrr^{(sb)}$,
\begin{eqnarray}
& &
\massrr^{(sb)} = R^\dagger\bmatrix{cc} \widetilde{m}_1^2 & 0
\\ 0 & \widetilde{m}_2^2 \ematrix R,
\nonumber \\
& &R\equiv \bmatrix{cc} \cos\theta & -\sin\theta e^{-i\sigma} \\
\sin\theta & \cos\theta e^{-i\sigma} \ematrix\;,
\end{eqnarray}
where $R$ absorbs the phase $\sigma$ and transfers it to the
quark-squark-gluino coupling, and $\theta$ is a measure of the
relative weight of $\tilde m^2_{23}$ and $(\tilde m^2_{22}-\tilde
m^2_{33})$. Clearly, with $\tilde m^2_{23}\sim \tilde
m^2_{22,33}$, in general we have one suppressed eigenvalue
$\widetilde m^2_1$ due to level splitting.

Our scenario corresponds to $\tilde m^2_{22}\simeq\tilde
m^2_{33}\simeq\tilde m^2_{23}\simeq\tilde m^2$ such that a
democratic structure is realized for right-handed squark mass
matrix in Eq.~(\ref{eq:sbbase}). The eigenstates hence carry both
$s$ and $b$ flavors and are called the strange-beauty squarks
$\widetilde{sb}_{1,2}$. To achieve this, some fine tuning is
necessary.
As a typical case, we take $\tilde m^2_{22}= \tilde
m^2_{33}=\tilde m^2$ ({\it i.e.} $\theta=\pi/4$ and $\tilde m^2_1+
\tilde m^2_2=2\tilde m^2$) and consider the ratio $\tilde
m^2_{23}/\tilde m^2 \equiv 1-\delta$. For small $\delta$, we have
$\tilde m^2_1/\tilde m^2 \approx \delta$ and $\tilde m^2_2/\tilde
m^2 \approx 2-\delta$, and a low mass $\widetilde{sb}_1$ can be
realized.
How light can $\widetilde m_1$ be depends on the fine tuning one
is willing to make. For $\tilde m=1\,(2)$ TeV, to get $\tilde
m_1\simeq 0.2$ TeV the level of tuning is
$\lambda^2\,(\lambda^3)$, which is comparable to what is seen in
$V_{\rm CKM}$. As had been shown in \cite{CHNphik},
${\cal S}_{\phi K_S}<0$ requires
$\widetilde{sb}_1$ to be suitably light, and the
gluino should not be too heavy. Since we would like to focus on
the effects from a light right-handed squark, we fix $\widetilde
m_1=0.2$ TeV in the following analysis. We will illustrate the
$\tilde m$ dependence of our results instead.

In our computations, we use mass eigenbasis of
Eq.~(\ref{eq:sbbase}) for right-handed squarks, since off-diagonal
elements are large. However, $(\widetilde M^2_d)^{sb}_{LR,RL}$
itself is strongly suppressed by down-type quark masses, and
off-diagonal elements of $\massll^{sb}$ are $\lambda^2$
suppressed. Hence we also use mass insertion approximation for
contributions arising from $q\,\tilde q_L\tilde g$, since
perturbative expansion is possible.

\section{\label{sec:result}Results for $B_s$}

$B_s$ mixing can be studied by flavor specific decays such as
$B_s\to D_s \pi^-$, but we are more interested in CP violating
measurables, especially those that could shed light on potential
new physics as hinted by ${\cal S}_{\phi K_S} < 0$ in $B_d$ system. In
the following, we give results on $B_s\to J/\psi\phi$, $K^+K^-$
and $\phi\gamma$ as they illustrate different aspects.
Since new physics effect on these decays is our interest,
we employ the naive factorization approach for calculation of
hadronic matrix elements.

%% discuss about Jpsi phi
\subsection{$B_s\to J/\psi\phi$}

\begin{figure}[t]
\vskip 3mm
\includegraphics[width=1.5in,height=1.3in,angle=0]{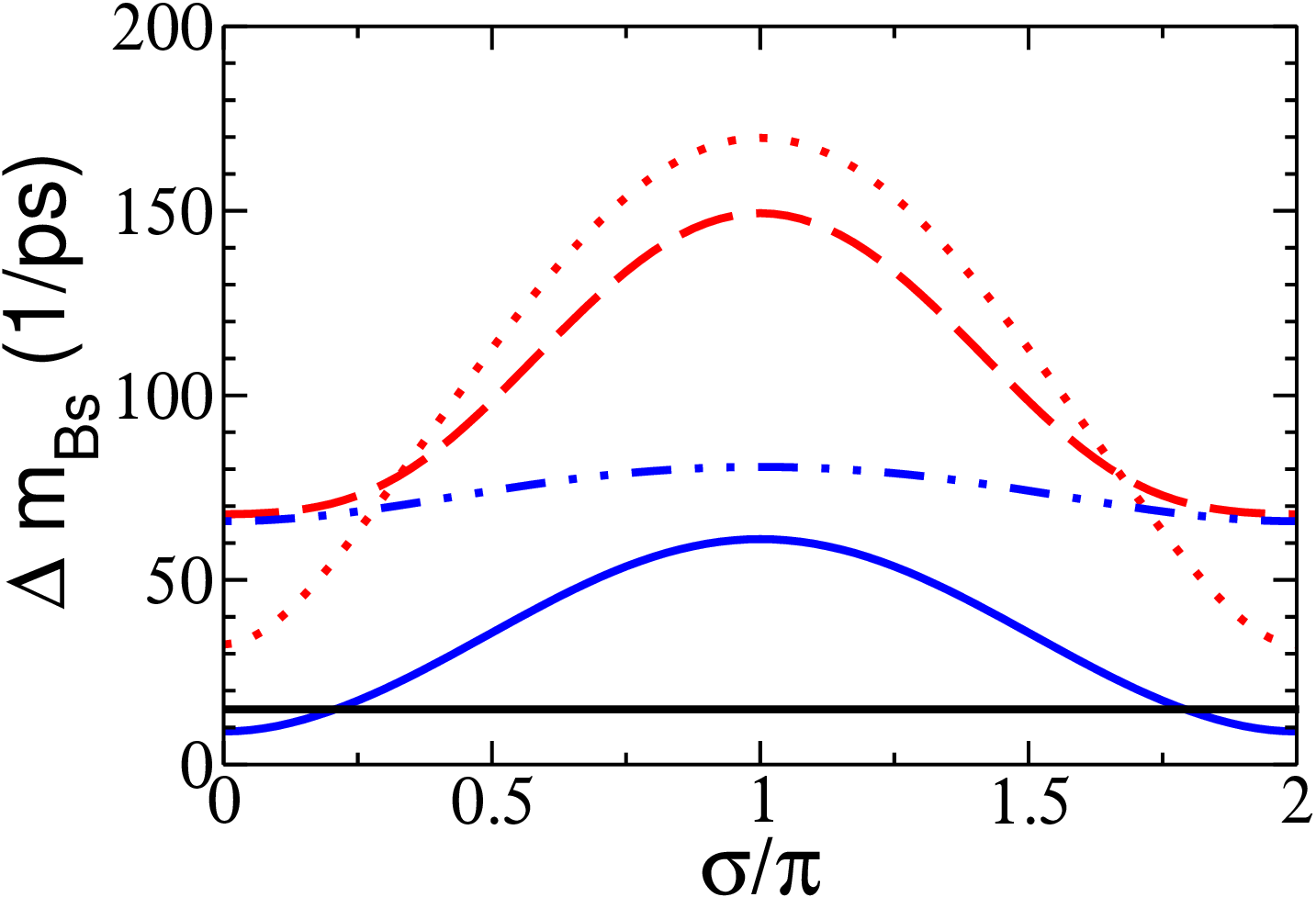}
\hspace{6.6mm}
\includegraphics[width=1.35in,height=1.3in,angle=0]{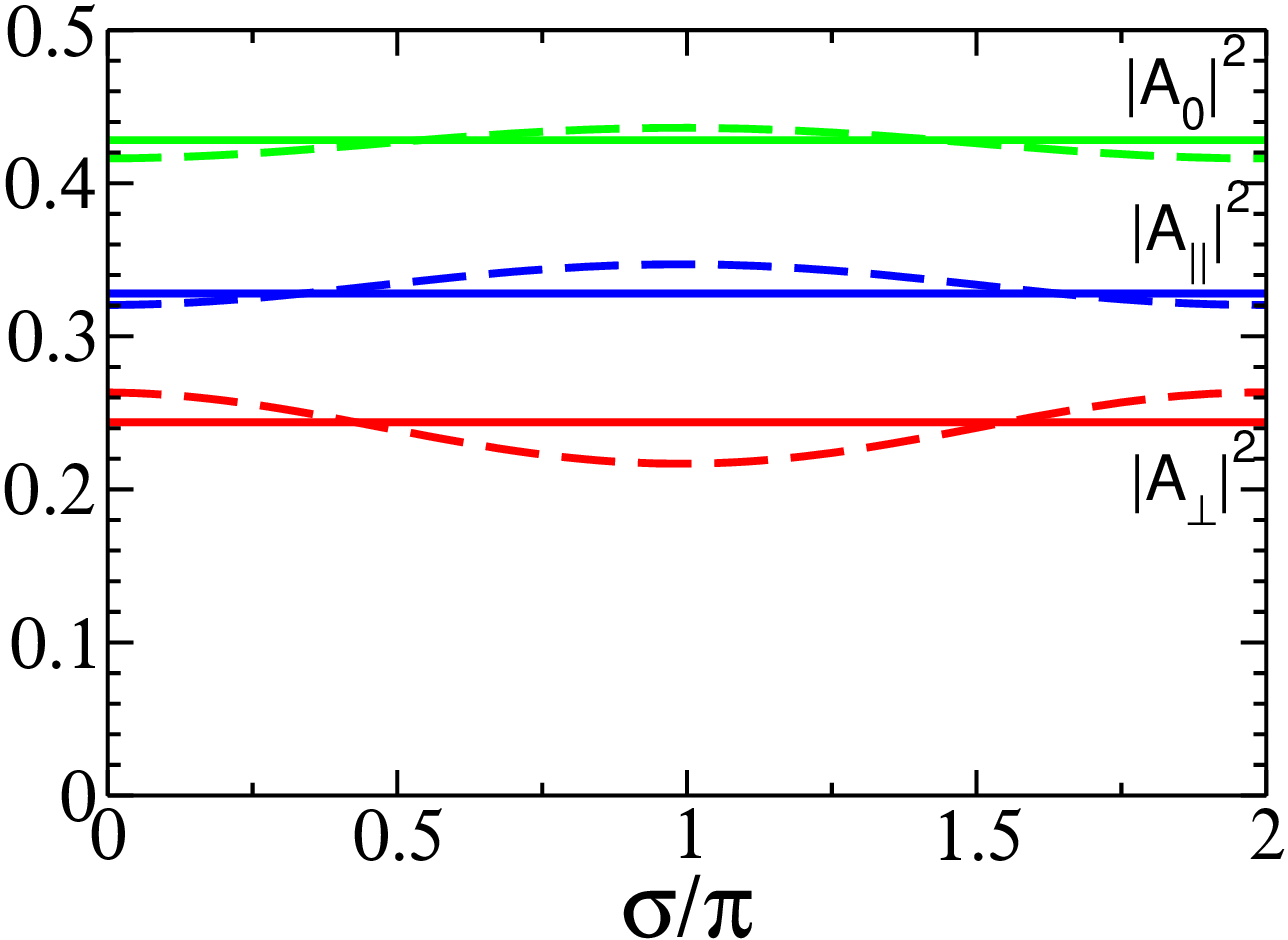}
\vskip 7mm
\includegraphics[width=1.5in,height=1.3in,angle=0]{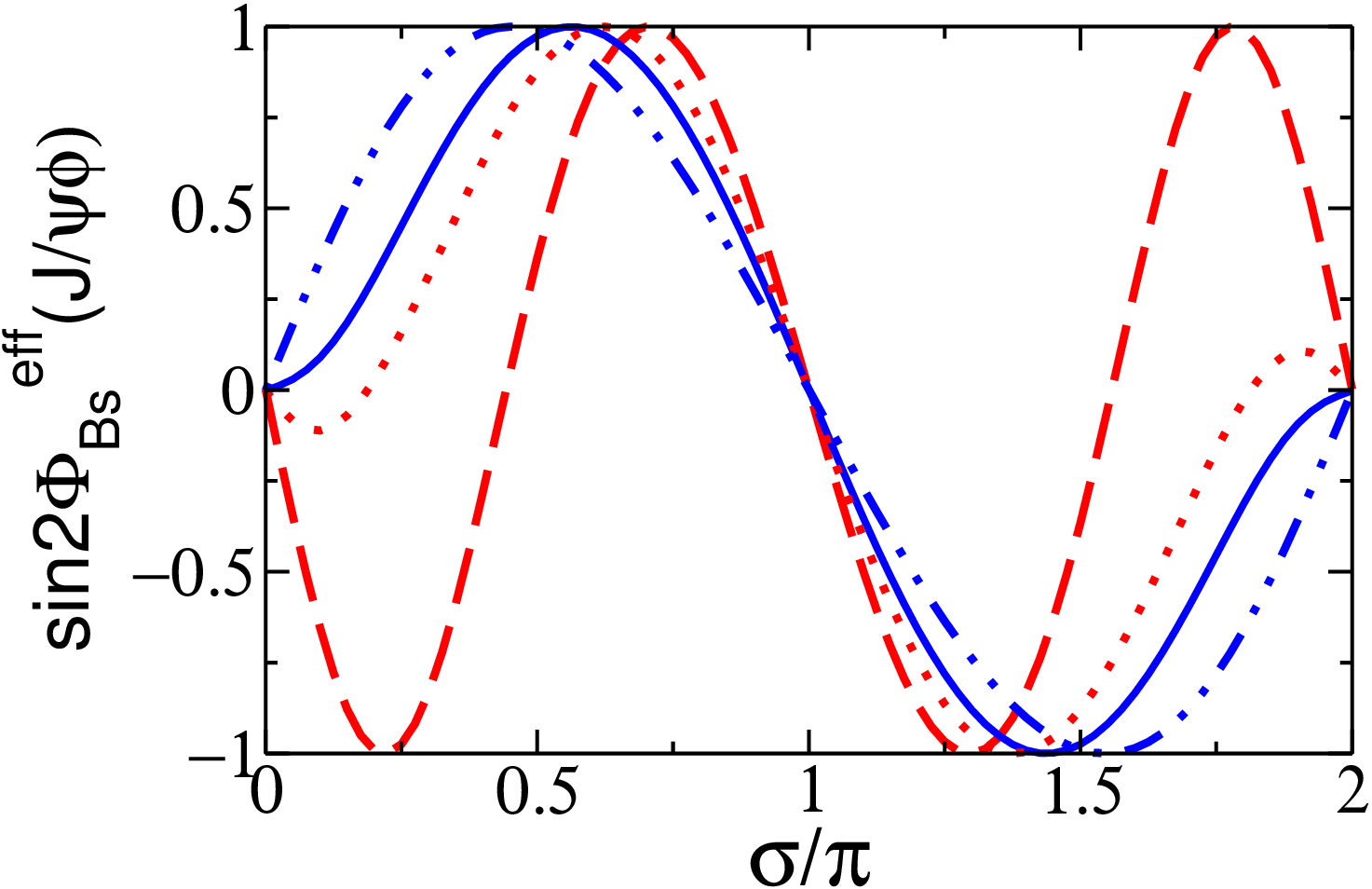}
\hspace{3mm}
\includegraphics[width=1.5in,height=1.3in,angle=0]{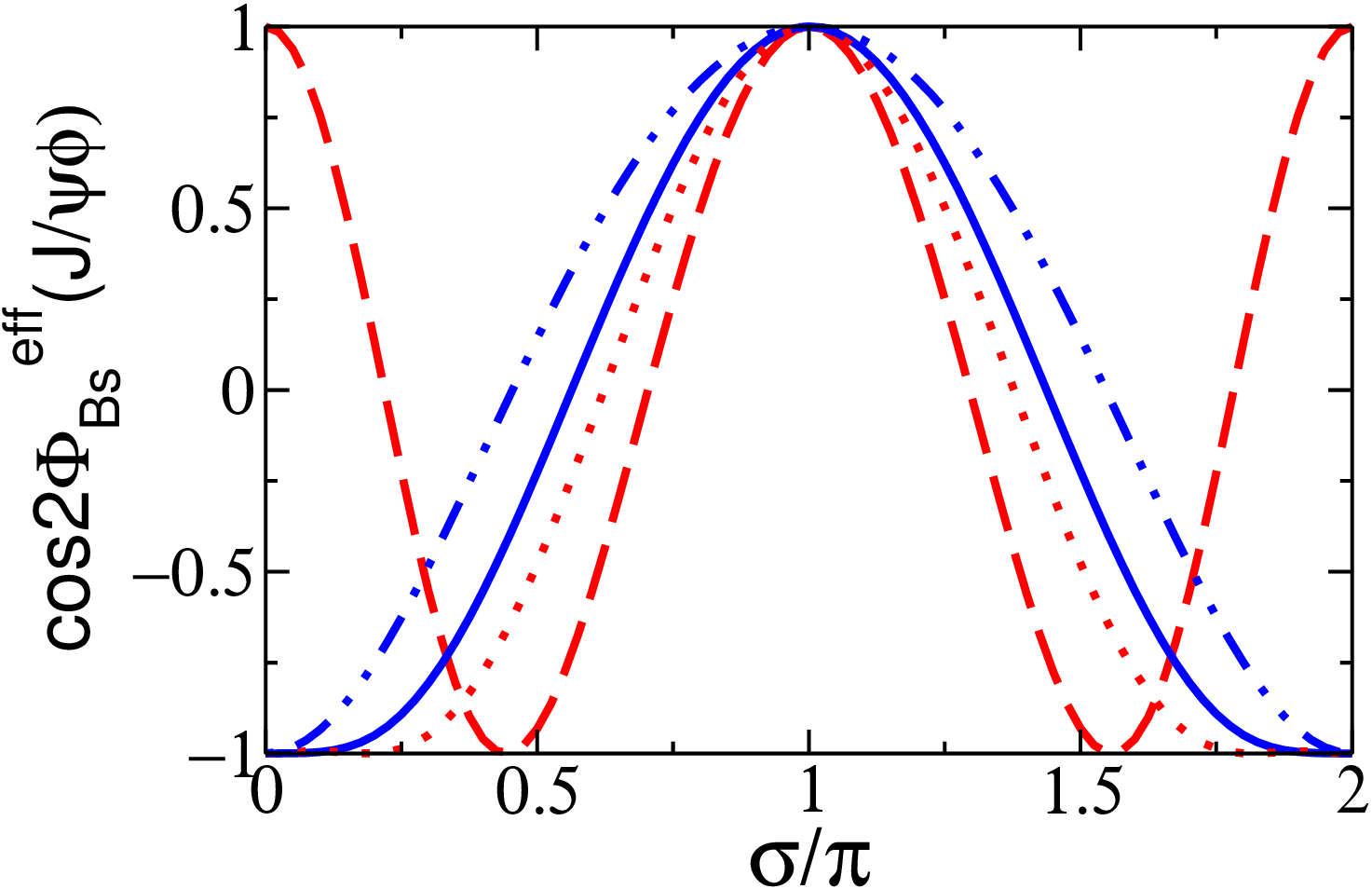}

\vskip 3mm \caption{{\label{fig:Psiphi}} (a) $\Delta m_{B_s}$, (b)
amplitudes $\vert A_i\vert^2$, (c)~$\sin 2\Phi_{B_s}^{\rm
eff}(J/\psi\phi)$ and (d) $\cos 2\Phi_{B_s}^{\rm eff}(J/\psi\phi)$
vs CP phase $\sigma$ (normalized to $\pi$) for $\tilde m_1=0.2$
TeV. For (a), (c) and (d), solid, dot-dash (dash, dots) lines are
for $\tilde{m} =$ 2, 1 TeV, $m_{\tilde g}=0.8\ (0.5)$ TeV. For
(b), solid line (and horizontal line in (a)) is the SM
expectation, while dash line is for $\{m_{\tilde g},\tilde
m\}=\{0.5,2\}$ TeV.
}
\end{figure}

The decay $B_s\to J/\psi\phi$ is a clean mode for the same reason
as $B_d\to J/\psi K_S$, {\it i.e.} a single tree amplitude
dominates and the relevant CKM matrix element $\cvcbcs$ is real.
Therefore, this decay mode is promising for the extraction of weak
phase information related to $B^0_s$-$\bar B^0_s$ mixing,
$\Phi_{B_s}$. Since the weak phase of $B^0_s$-$\bar B^0_s$ mixing
is basically absent in SM, the values of $\sin 2\Phi_{B_s}$ and
$\cos 2\Phi_{B_s}$ are governed by new physics effects. Assuming
$(q/p)\equiv e^{2i\Phi_{B_s}}$ with ${\rm Re}(q/p)$ positive in
sign within the SM, we have
\begin{eqnarray}
\sin 2\Phi_{B_s}\equiv
\sin 2(\Phi_{B_s}^{\rm SM}+ \Phi_{B_s}^{\rm NP})=\sin 2\Phi_{B_s}^{\rm NP},
\nonumber \\
\cos 2\Phi_{B_s}\equiv
\cos 2(\Phi_{B_s}^{\rm SM}+ \Phi_{B_s}^{\rm NP})=\cos 2\Phi_{B_s}^{\rm NP},
\label{eq:PhiBs}
\end{eqnarray}
to good accuracy. Fig.~\ref{fig:Psiphi}(a) shows our predictions
for $\Delta m_{B_s}$. As mentioned in Introduction, the large
effect of ${\cal S}_{\phi K_S}<0$ implies $\Delta m_{B_s}\gtrsim
70$ ${\rm ps}^{-1}$. The measurement of $\sin 2\Phi_{B_s}$ would
be challenging, but it clearly is a good probe of new physics.
Furthermore, if $\Delta\Gamma_s/\Gamma_s$ is around $10\%$, it
could also be available for the study of CP violation due to $\cos
2\Phi_{B_s}$. Let us see how these two quantities can be measured
via $B_s\to J/\psi\phi$ decay.

In contrast to $B_d\to J/\psi K_S$, the decay product of $B_s\to
J/\psi\phi$ consists of two vector mesons, and is not a CP
eigenstate. However, angular analysis of $B_s\to
J/\psi(l^+l^-)\phi(K^+K^-)$ can distinguish between the CP-even
and CP-odd components of the full amplitude, with 6 measurable
components associated with the angles describing the kinematics
\cite{Dighe,London}. The current experimental result for the
branching fraction is $(9.3\pm 3.3)\times 10^{-4}$ \cite{PDG}.
Moreover, the CDF collaboration finds $|A_0|^2\approx 0.61\pm 0.14$ and
$|A_\perp|^2\approx 0.23\pm 0.19$ \cite{CDF}, with
$|A_0|^2+|A_\parallel|^2+|A_\perp|^2=1$.

%
% PLOT
%

We write the full decay amplitude as follows:
\begin{eqnarray}
& &A(B^0\to f)=
A_0\,g_0+A_{\parallel}\,g_{\parallel}+iA_{\perp}\,g_{\perp},
\nonumber \\
& &\bar A(\bar B^0\to f)=
\bar A_0\,g_0+\bar A_{\parallel}\,g_{\parallel}-i\bar A_{\perp}\,g_{\perp},
\end{eqnarray}
where 
$g_{0(\parallel,\perp)}$ depends on kinematic angles.
For $B_s\to J/\psi\phi$, the three polarization amplitudes $A_0$,
$A_\parallel$ (CP-even) and $A_\perp$ (CP-odd) are expressed as
\cite{Dighe},
\begin{eqnarray}
& & \frac{A_0}{A_\parallel}=-\frac{1}{\sqrt 2}\;
\biggl[x - \frac{2m_{J/\psi}m_\phi}{(m_{B_s}+m_\phi)^2}\frac{A_2}{A_1}
\frac{C^b_{J/\psi\phi}}{C^a_{J/\psi\phi}}(x^2-1)\biggr],
\nonumber \\
& & \frac{A_\perp}{A_\parallel}=
\frac{2m_{J/\psi}m_\phi}{(m_{B_s}+m_\phi)^2}
\frac{V\; C^c_{J/\psi\phi}}{A_1\; C^a_{J/\psi\phi}}\sqrt{x^2-1},
\label{eq:polaraio}
\end{eqnarray}
where $x\equiv p_{J/\psi}\cdot p_\phi/(m_{J/\psi} m_\phi)$,
and are sensitive to the form factors $A_1(m_{J/\psi}^2)$,
$A_2(m_{J/\psi}^2)$ and $V(m_{J/\psi}^2)$. In the SM the short
distance coefficients $C^{a(b,c)}_{J/\psi\phi}$ are given by,
\begin{eqnarray}
C^{a(b,c)}_{J/\psi\phi}&=&\sqrt{2}G_F\cvcbcs f_{J/\psi}m_{J/\psi}\nonumber \\
& & \hspace{1cm} \times
\biggl[a_2-\frac{\cvtbts}{\cvcbcs}(a_3+a_5)\biggr],
\end{eqnarray}
where, for example, $a_2$ is the coefficient of the $O_2=(\bar
s_\alpha c_\beta)_{V-A}(\bar c_\alpha b_\beta)_{V-A}$ operator. In
addition to the SM contributions, we take into account the SUSY
contributions due to $q$-$\tilde g$-$\tilde q$ interaction,
including the contributions coming from chromo-dipole operator
associated with $\bar sbg$~\cite{Soddu}. Such new effects in
general can provide differences between $C^{a(b,c)}_{f}$.

Here we use
$\{A_1(m_{J/\psi}^2),A_2(m_{J/\psi}^2),V(m_{J/\psi}^2)\}=\{0.42,0.47,0.87\}$
based on the central values at $q^2=0$ in the light-cone sum rule
(LCSR) approach~\cite{Ball}. The theoretical uncertainty is
estimated as 15\%. Fig.~\ref{fig:Psiphi}(b) illustrates the
polarization amplitudes. With these form factors, there is
$1.3\;\sigma$ deviation from the measurement $|A_0|^2\sim 0.61$,
which is acceptable. It may be experimental, or may be due to
sensitivity of form factors. Since this decay is tree dominant,
the new physics effect is rather tiny. For example, taking
$\{\tilde m_1,m_{\tilde g},\tilde m\}=\{0.2,0.5,2\}$ TeV, the
impact is less than 10\% (dash line in Fig.~\ref{fig:Psiphi}(b)).
Despite ambiguities from nonperturbative effects, we find the
dominant component, with or without NP effect, is the CP-even
state $(|A_0|^2+|A_\parallel|^2)$. In particular, the longitudinal
component {\it i.e.} $|A_0|^2$ is found to be dominant.

The absolute branching fraction is sensitive to form factors and
also the effective number of colors, $N_c^{\rm eff}$. This is
because the decay is color-suppressed. For SM expectation, we find
${\cal B}_{\rm SM}\sim 7.0\;(9.4,\; 12.3)\;\times 10^{-4}$ for
$N_c^{\rm eff}=2.3\;(2.2,\;2.1)$, while the experimental
measurement is ${\cal B}\sim (6\sim 13)\;\times 10^{-4}$.
Throughout this paper, we take the effective number of color to be
$N_c^{\rm eff}=2.3$, giving $a_2\sim 0.14$.

%%%%
The direct CP violation asymmetries $A_{J/\psi\phi}$ for each
polarization amplitude is
\begin{eqnarray}
& &{\cal A}_f=\frac{|\bar A_\lambda|^2-|A_\lambda|^2} {|\bar
A_\lambda|^2+|A_\lambda|^2}, \hspace{7mm} (\lambda=0,\;
\parallel,\; \perp),
\end{eqnarray}
which are negligibly small because of single tree dominance and
absence of absorptive parts (assuming absence of final state
interaction phases).
Thus, the actual CP probes are ${\cal S}_{J/\psi\phi}$ and
${\cal A}_{\Delta\Gamma}$ of Eq.~(\ref{eq:timeCPII}), which should
satisfy
$\vert {\cal }S_{J/\psi\phi}\vert^2 +
\vert{\cal A}_{\Delta\Gamma}\vert^2 \cong 1$,
with equality cross-checked by
direct CP violation.
This allows us to introduce the terminology of $\sin
2\Phi_{B_s}^{\rm eff}(J/\psi\phi)$ and $\cos 2\Phi_{B_s}^{\rm
eff}(J/\psi\phi)$ instead of ${\cal S}_{J/\psi\phi}$ and ${\cal
A}_{\Delta\Gamma}$, so we can compare with the CP violating phase
induced purely by $B^0_s$-$\bar B^0_s$ mixing, $\sin 2\Phi_{B_s}$
and $\cos 2\Phi_{B_s}$. Let us concentrate on the dominant
longitudinal component $|A_0|^2$.
We find that, to good approximation,
\begin{eqnarray}
& & \sin 2\Phi_{B_s}^{\rm eff}(J/\psi\phi)\approx \sin 2\Phi_{B_s},
\nonumber \\
& & \cos 2\Phi_{B_s}^{\rm eff}(J/\psi\phi)\approx \cos 2\Phi_{B_s},
\end{eqnarray}
because of $A_{J/\psi\phi}\approx 0$ and $A_0\approx +\bar A_0$.
The same result is obtained for measurements done with
$A_\parallel$, but for $A_\perp$ there is a sign change.
Thus, the weak phase of $B^0_s$-$\bar B^0_s$ mixing can be
measured via CP violation studies in the decay $B_s\to
J/\psi\phi$, which we illustrate for our model in
Figs.~\ref{fig:Psiphi}(c) and (d) vs the CP phase $\sigma$.

Our results for  $\sigma\sim 65^\circ$ are clearly rather
different from $\sin 2\Phi_{B_s}^{\rm eff}(J/\psi\phi) \sim 0$ and
$\cos 2\Phi_{B_s}^{\rm eff}(J/\psi\phi)\sim 1$ predicted by SM. As
noted~\cite{CHNphik}, for $\sigma\sim 65^\circ$, $\sin
2\Phi_{B_s}^{\rm eff}(J/\psi\phi)$ can vary over a rather wide
range, so a precision measurement could help pin down $\sigma$. It
is remarkable that the large effect of ${\cal S}_{\phi K_S}<0$ implies
$\cos 2\Phi_{B_s}^{\rm eff}(J/\psi\phi)<0$, that is, $\cos
2\Phi_{B_s}<0$. The measurements of $\sin 2\Phi_{B_s}^{\rm eff}$
and $\cos 2\Phi_{B_s}^{\rm eff}$ can shed light on our model
parameters.

We stress that $\cos 2\Phi_{B_s}^{\rm eff}$ is actually measured
via untagged $B_s$ data utilizing the lifetime difference between
two $B_s$ mass eigenstates, i.e. Eq.~(\ref{eq:untagg}). Even if
$\sin 2\Phi_{B_s}^{\rm eff}(J/\psi\phi)$ measurement gets hampered
by very fast $B_s$ oscillations, measurement of $\cos
2\Phi_{B_s}^{\rm eff}$ is in principle possible, so long that
$\Delta\Gamma_s$ itself can be measured.

%% discuss about K K
\subsection{$B_s\to K^+K^-$}

$B_s\to K^+K^-$ decay is dominated by QCD penguins and is similar
to $B_d\to K^+\pi^-$, except for the advantage that the final
state $K^+K^-$ is a CP (even) eigenstate. The decay amplitude is
written as
% amplitude in
\begin{eqnarray}
 iA(\bar B_s & \to & K^+K^-) = %\nonumber \\ & &
\frac{G_F}{\sqrt 2}f_K F_{0}^{B_sK}(m_{K}^2)\,
(m_{B}^2-m_{K}^2) \nonumber \\
 & \times & \Biggl\{ \vubus \, a_1
- \vtbts \;\Biggl[
\Delta a_4 + a_{10} \nonumber \\
& + & (\Delta a_6+a_8)R_P + \Delta c_{12}
\frac{\alpha_s}{4\pi}\frac{m_b^2}{q^2} \tilde S_{KK} \Biggr]
\Biggr\}\;, \label{eq:ampKK}
\end{eqnarray}
where $f_K$ and $F_{0}^{B_{s}K}$ are decay constant and form
factor of $B_{s(d)}\to K$ transition, respectively, and $R_P$ is a
chiral enhancement factor, where we use $R_P=1.24$. We write
$\Delta a_i= a_i-a^\prime_i$ and $\Delta c_{12}=
c_{12}-c^\prime_{12}$, where the coefficients $a^\prime_i$ and
$c^\prime_{12}$ are related to the NP right-handed dynamics. The
CP conserving phases are taken into account via BSS mechanism
({\it e.g.} see~\cite{YHCHEN}). The last term in Eq.
(\ref{eq:ampKK}) is induced by chromo-dipole operator, and
accompanied by the hadronic factor $\tilde S_{KK}/q^2$. We shall
use~\cite{CHNphik} $\tilde S_{KK}=-1.58$ as evaluated from naive
factorization, and $q^2=m_b^2/3$ for the virtual gluon momentum
emitted by the $b\to s$ chromo-dipole.

\begin{figure}[t]
\includegraphics[width=1.5in,height=1.3in,angle=0]{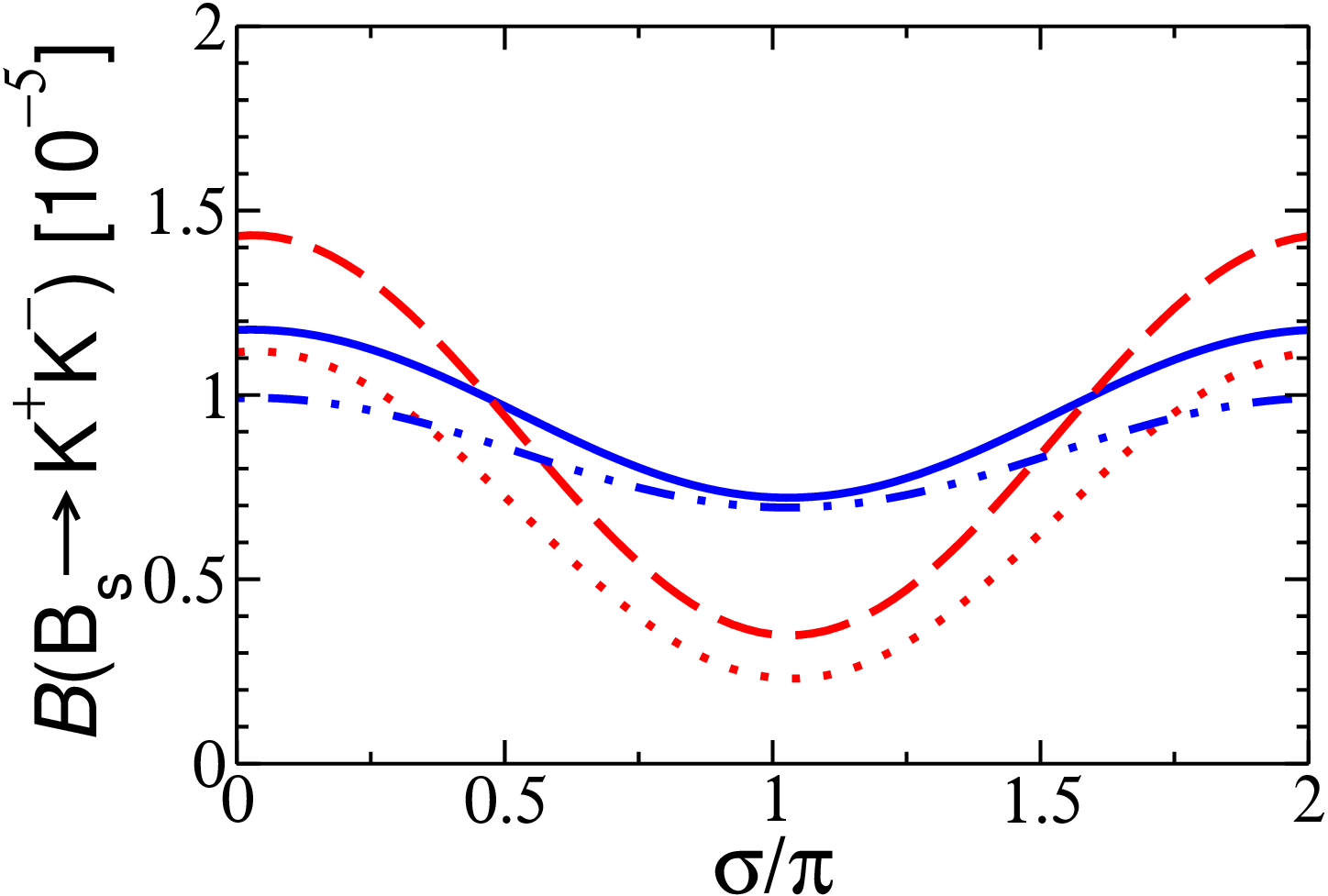}
\hspace{3mm}
\includegraphics[width=1.5in,height=1.3in,angle=0]{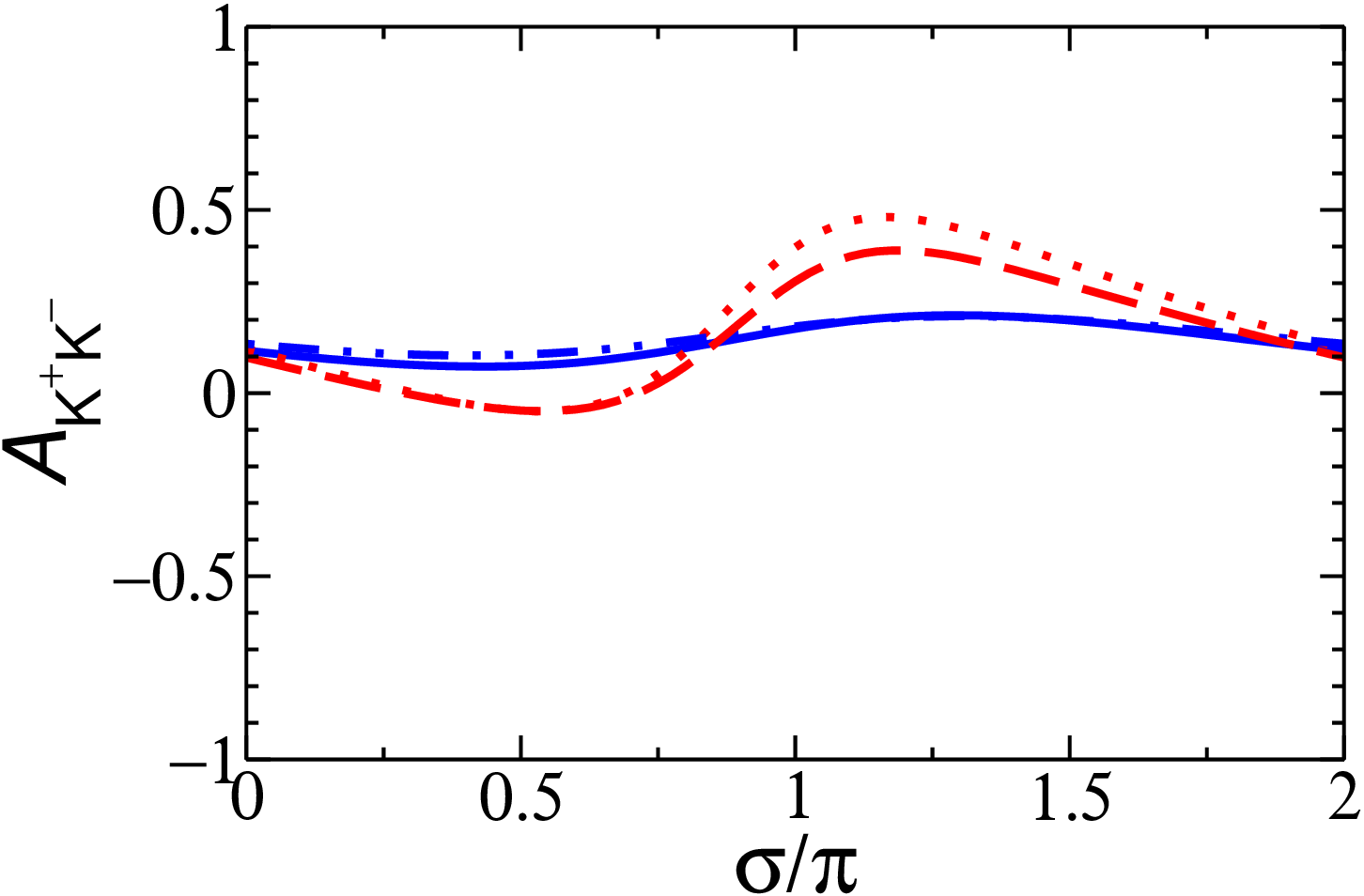}

\vskip 7mm
\includegraphics[width=1.5in,height=1.3in,angle=0]{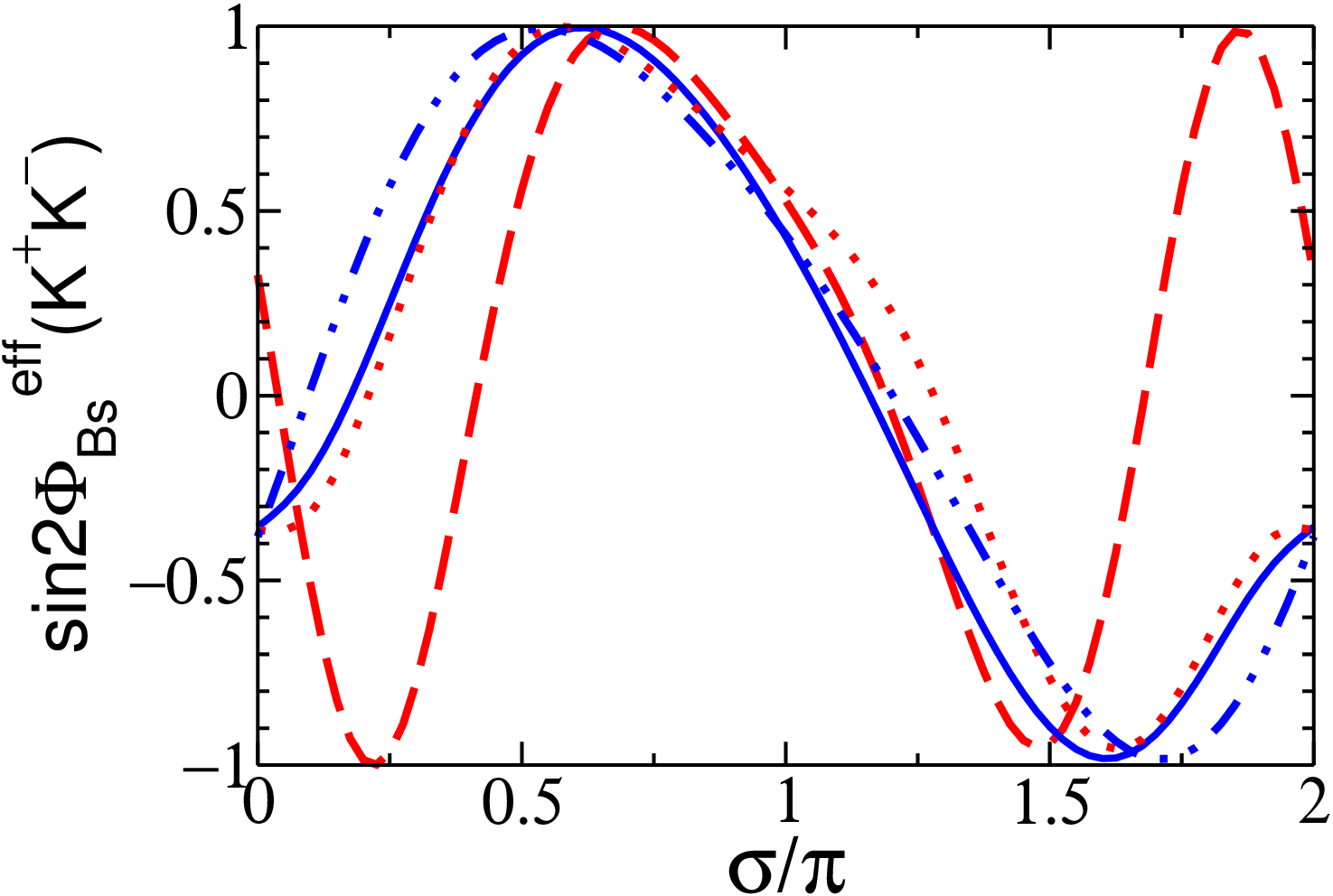}
\hspace{3mm}
\includegraphics[width=1.5in,height=1.3in,angle=0]{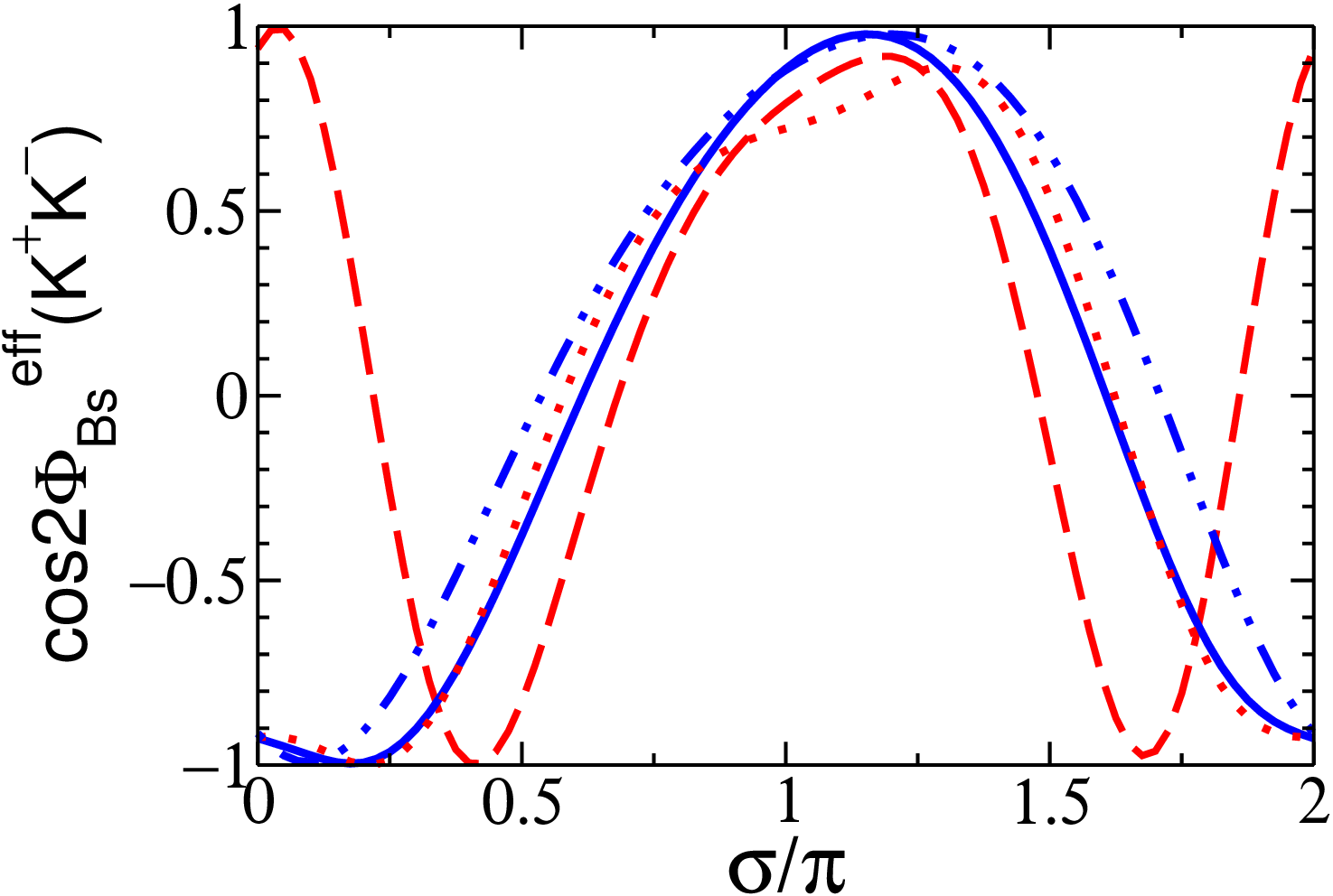}

\vskip 7mm
\includegraphics[width=1.5in,height=1.3in,angle=0]{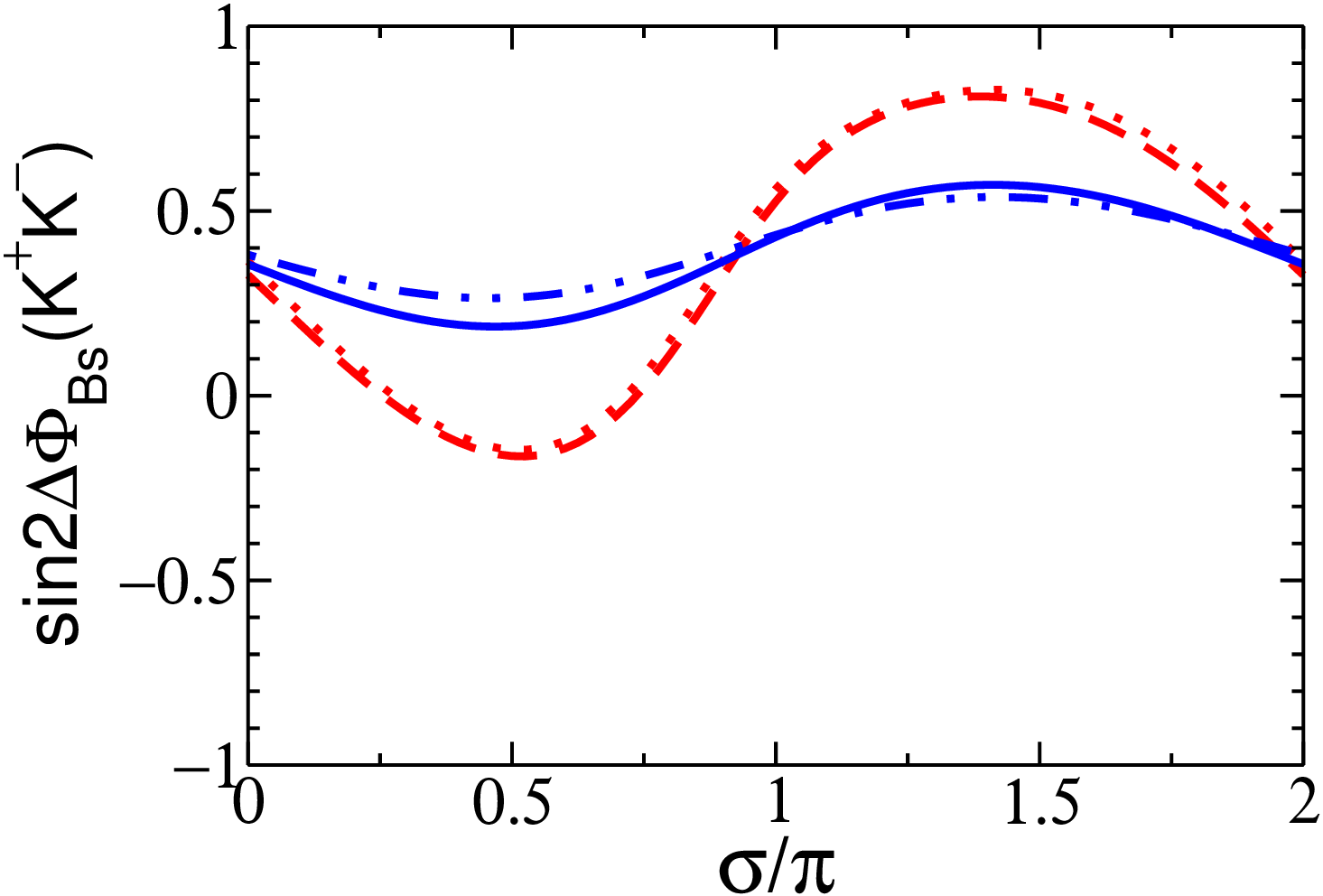}
\hspace{3mm}
\includegraphics[width=1.5in,height=1.3in,angle=0]{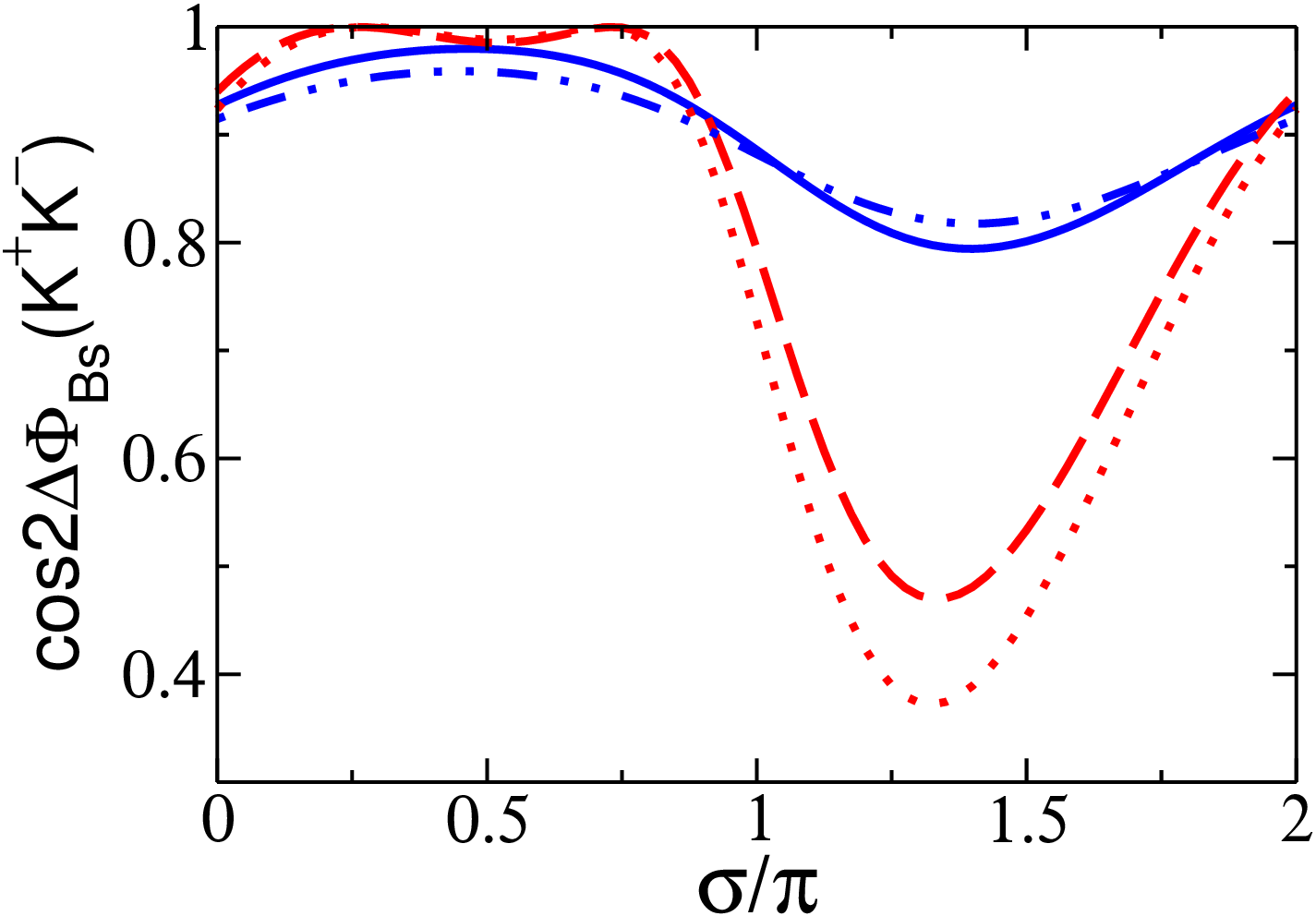}
\vskip 5mm \caption{{\label{fig:bskk}} For $B_s\to K^+K^-$
 (a) branching fraction (see text),
 (b) $A_{K^+K^-}$, (c) $\sin 2\Phi_{B_s}^{\rm eff}$,
 (d) $\cos 2\Phi_{B_s}^{\rm eff}$,
 (e) $\sin 2\Delta\Phi_{B_s}$ and
 (f) $\cos 2\Delta\Phi_{B_s}$ vs $\sigma$, with notation as in
Fig.~\ref{fig:Psiphi}(a).
%Solid vertical line indicates $\sigma\sim 65^\circ$.
}
\end{figure}

Eq. (\ref{eq:ampKK}) illustrates several problems for extracting
CP violating phases from $B_s\to K^+K^-$. In contrast to the clean
$B_s\to J/\psi\phi$ mode, the tree contribution is suppressed by
$\vubus$ and penguin contributions dominate, hence the amplitude
is sensitive to hadronic parameters. Furthermore, the tree
contribution brings in the CKM phase $\phi_3 \equiv \arg
V_{ub}^*$. We note, however, that the latter can be in principle
extracted from $B_s\to D_s^\pm K^\mp$ decays, independent from the
analogous $B_d\to D^\pm K^\mp$ program. The latter should still
provide us with information on $\phi_3$ if fast $B_s$ oscillations
degrade our sensitivity in the $B_s\to D_s^\pm K^\mp$ program.
At the moment, especially in presence of new physics, it is
not clear which $\phi_3$ value to take. We take $\phi_3=60^\circ$ in the
following.

The branching fraction is sensitive to the form factor $F_{0}^{B_sK}$.
Figure~\ref{fig:bskk}(a) shows our result of the branching fraction,
with $F_{0}^{B_sK}=F_{0}^{B_dK}=0.33$.
The current upper bound for the rate of $B_s\to K^+K^-$ is
\begin{eqnarray}
{\cal B}(B_s\to K^+K^-) < 5.9\times 10^{-5},
\end{eqnarray}
at $90\%$ confidence level \cite{PDG}. On the other hand, recent
results from the Tevatron~\cite{Tevatoron} give,
\begin{eqnarray}
\frac{f_{b\to B_s}}{f_{b\to B_d}} \, \frac{{\cal B}(B_s\to
K^+K^-)}{{\cal B}(B_d\to K^+\pi^-)} = 0.74\pm 0.20\pm 0.22,
\label{eq:CDFdata}
\end{eqnarray}
where $f_{b\to B_{s(d)}}$ is production fraction for $B_{s(d)}$,
and $f_{b\to B_s}/f_{b\to B_d} \approx 0.27$ \cite{PDG}. This
suggests ${\cal B}(B_s\to K^+K^-)\sim 3\times {\cal B}(B_d\to
K^+\pi^-)$, and seems to prefer $F_{0}^{B_sK}$ to be 1.5 to 1.8
times larger than the 0.33 value we use.

The quantities ${\cal A}_{K^+K^-}$, ${\cal S}_{K^+K^-}$ and ${\cal
A}_{\Delta\Gamma}$, which are independent of $F_{0}^{B_sK}$ but
correlated to each other, are illustrated in
Fig.~\ref{fig:bskk}(b)-(d).
Let us focus on the region $\sigma\sim 65^\circ$. Because ${\cal
A}_{K^+K^-}$ is tiny, we can still use the same terminology of
$\sin 2\Phi_{B_s}^{\rm eff}(K^+K^-)$ and $\cos 2\Phi_{B_s}^{\rm
eff}(K^+K^-)$. We see that the decay phase $\phi$, defined by
$A=|A|e^{i\phi}$, must be rather suppressed, since $\sin
2\Phi_{B_s}^{\rm eff}(K^+K^-)$ and $\cos 2\Phi_{B_s}^{\rm
eff}(K^+K^-)$ are quite similar to $B_s\to J/\psi\phi$ case, {\it
i.e.} $\sin 2\Phi_{B_s}$ and $\cos 2\Phi_{B_s}$, respectively.

For more detailed understanding of the decay phase $\phi$, we
define the difference angle $\Delta\Phi_{B_s}$ between
$\Phi_{B_s}^{\rm eff}$ and $\Phi_{B_s}$ which are given by (up to
discrete ambiguities)
\begin{eqnarray}
\sin 2\Delta\Phi_{B_s}
\equiv \sin 2(\Phi_{B_s}^{\rm eff}-\Phi_{B_s}),
\nonumber \\
\cos 2\Delta\Phi_{B_s}
\equiv \cos 2(\Phi_{B_s}^{\rm eff}-\Phi_{B_s}).
\label{eq:deltaPhi}
\end{eqnarray}
Since $\Phi_{B_s}$ is just $\Phi_{B_s}^{\rm eff}(J/\psi\phi)$ to
good approximation, one can extract $\Delta\Phi_{B_s}$ by using
$\Phi_{B_s}^{\rm eff}(J/\psi\phi)$ instead of $\Phi_{B_s}$. The SM
gives $\Delta\Phi_{B_s}^{\rm SM}\sim 10^\circ$.
Fig.~\ref{fig:bskk}(e) and (f) illustrate $\sin 2\Delta\Phi_{B_s}$
and $\cos 2\Delta\Phi_{B_s}$, where the vertical range for latter
is from 0.3 to 1 to reveal better detail. We see that for
$\sigma\sim 65^\circ$, $\sin 2\Delta\Phi_{B_s}$ has turned
negative for lower gluino mass case, and measurement can provide
some information. On the other hand, $\cos 2\Delta\Phi_{B_s} \cong
1$, and not much can be learned.

Although the measurement of these quantities may suffer from very
fast $B_s$ oscillations, our results for $S_{\phi K_S}<0$ imply
that $\Delta\Phi_{B_s}$ from $B_s\to K^+K^-$ decay can potentially
help determine model parameters.

%% discuss about phi gamma
\subsection{$B_s\to\phi\gamma$}

For radiative $b\to s\gamma$ transition, the decay rate at leading
order is proportional to $|c_{11}|^2$ and $|c_{11}^\prime|^2$,
%in the form of $\Gamma(B\to M^0\gamma) \propto |c^{(\prime)}_{11}|^2$.
where $c_{11}$, $c_{11}^\prime$ are the short-distance Wilson
coefficients of
\begin{eqnarray}
O_{11},\ O_{11}^\prime = \frac{e}{8\pi^2} m_b\bar
s\sigma^{\mu\nu}(1\pm\gamma_5)F_{\mu\nu}b.
\end{eqnarray}
In the SM with purely left-handed interaction, $c^\prime_{11}$ is
suppressed by $s$ quark mass hence negligible.

\begin{figure}[t]
\includegraphics[width=1.5in,height=1.3in,angle=0]{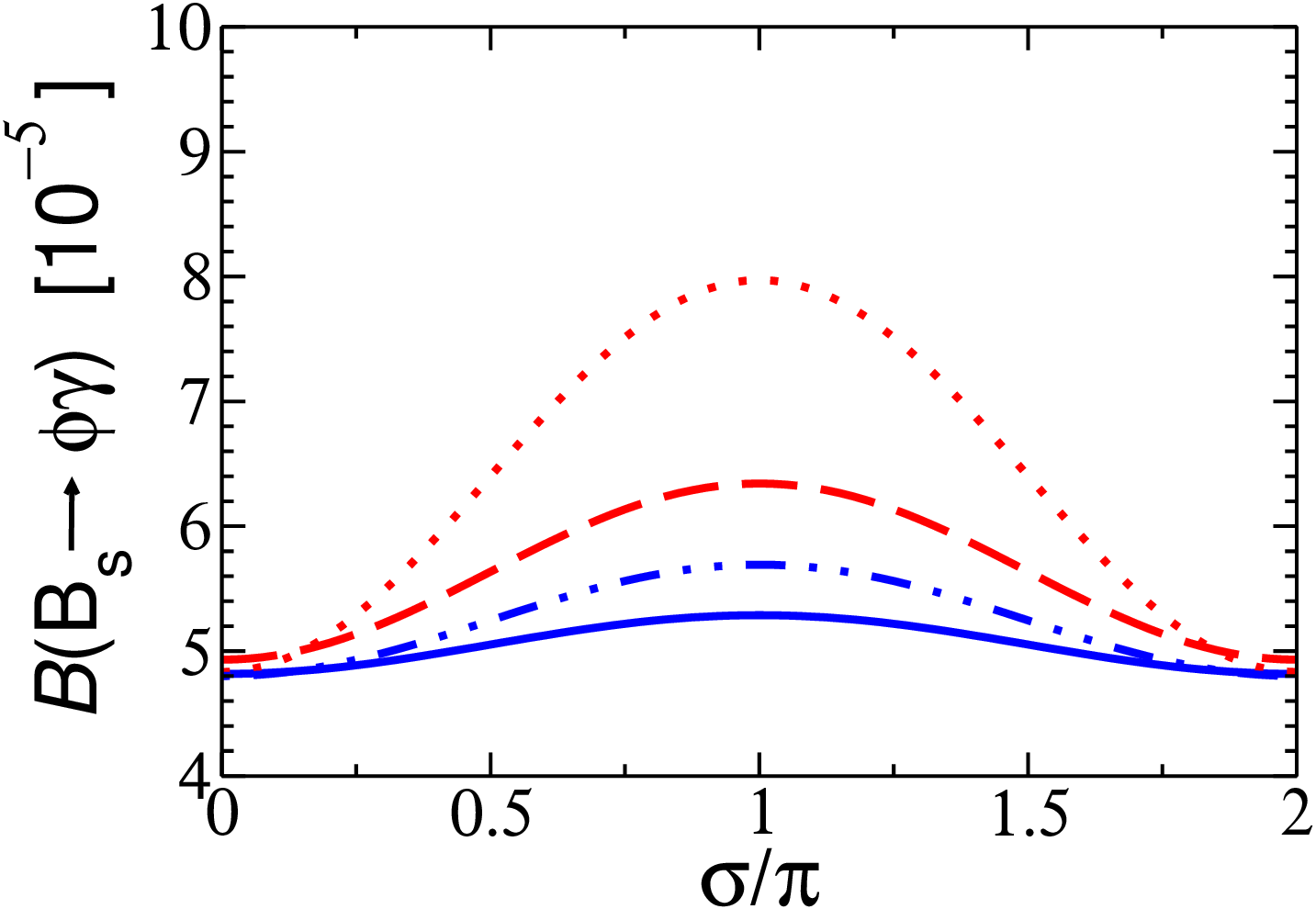}

\vskip 6mm
\includegraphics[width=1.5in,height=1.3in,angle=0]{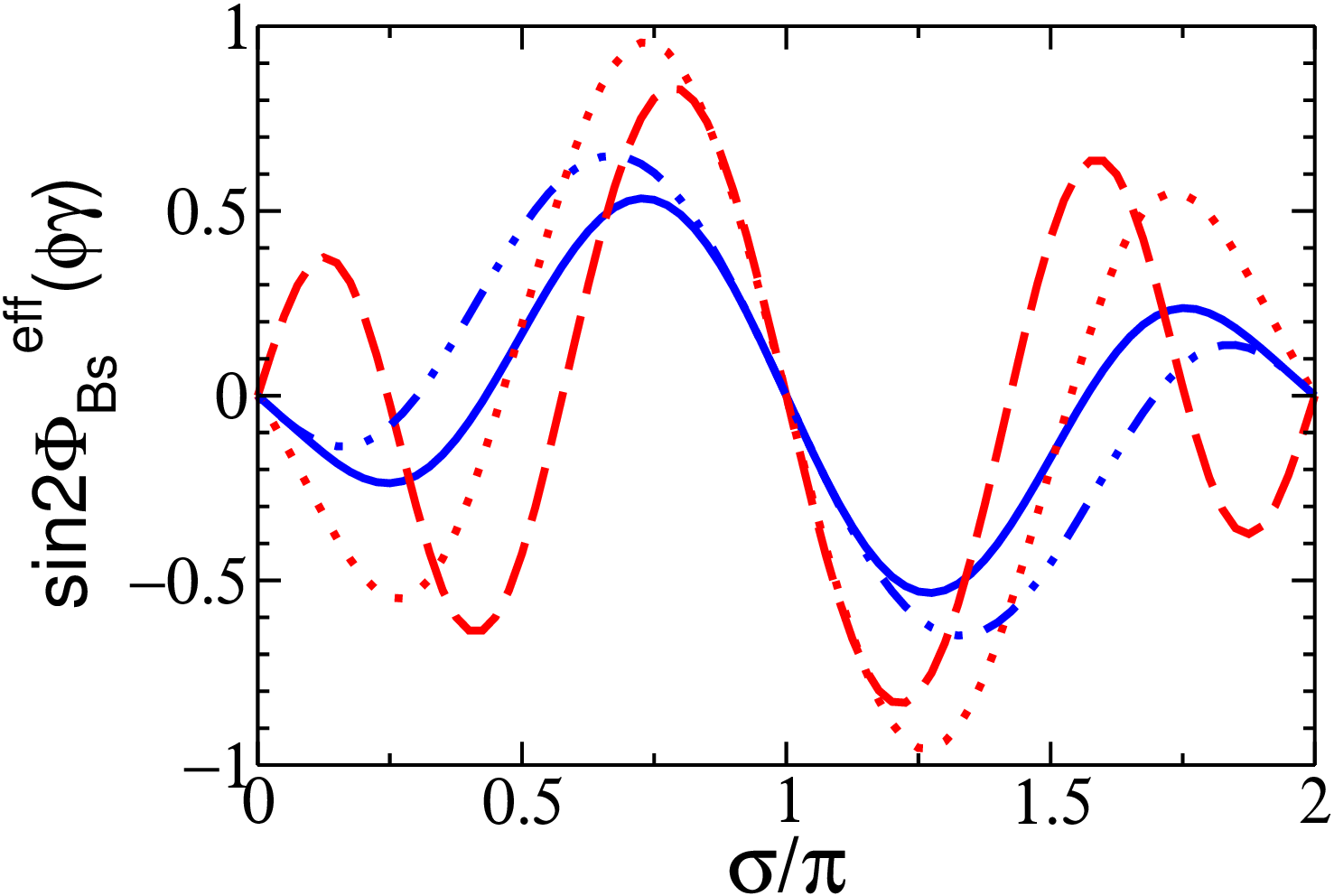}
\hspace{3mm}
\includegraphics[width=1.5in,height=1.3in,angle=0]{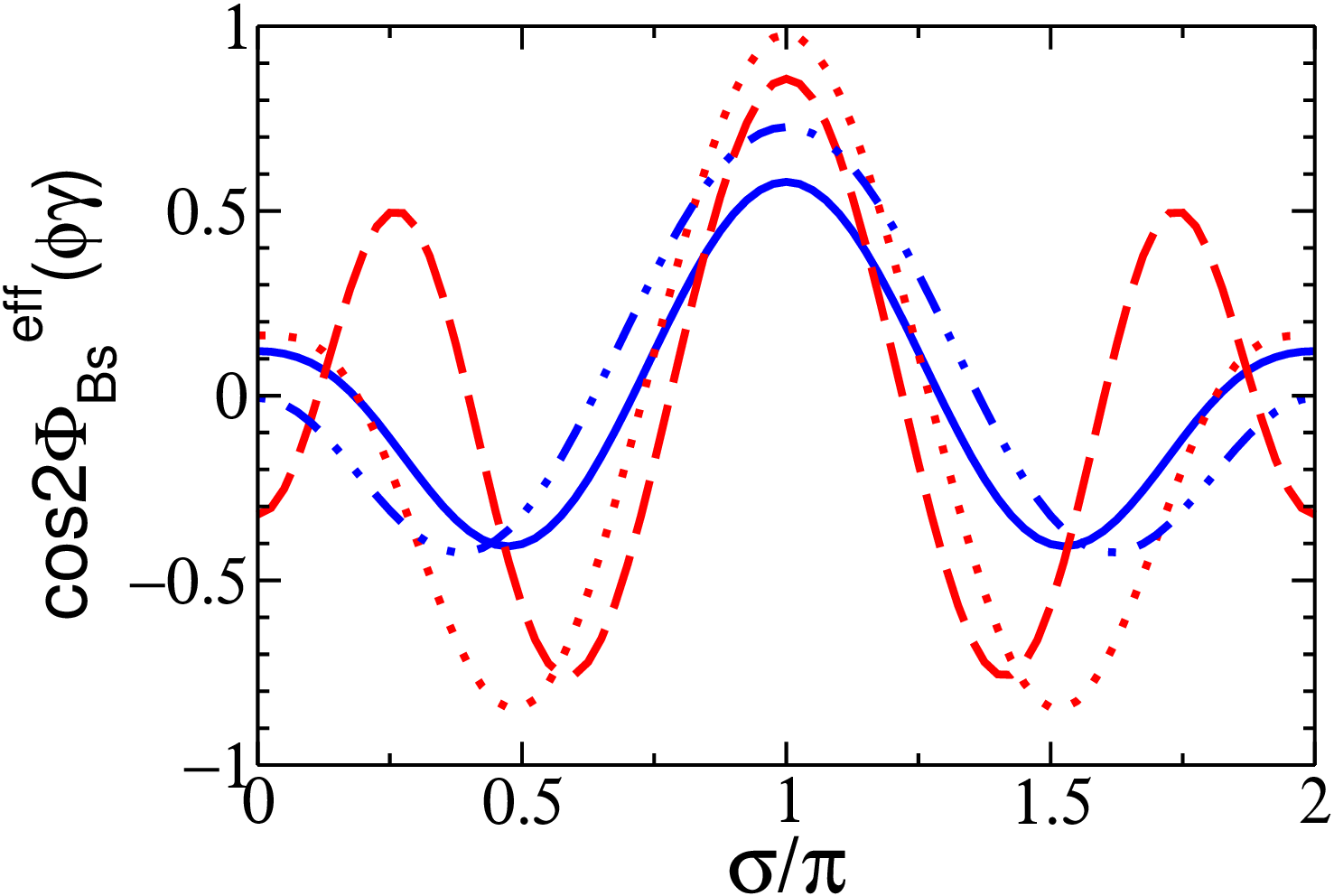}

\vskip 6mm
\includegraphics[width=1.5in,height=1.3in,angle=0]{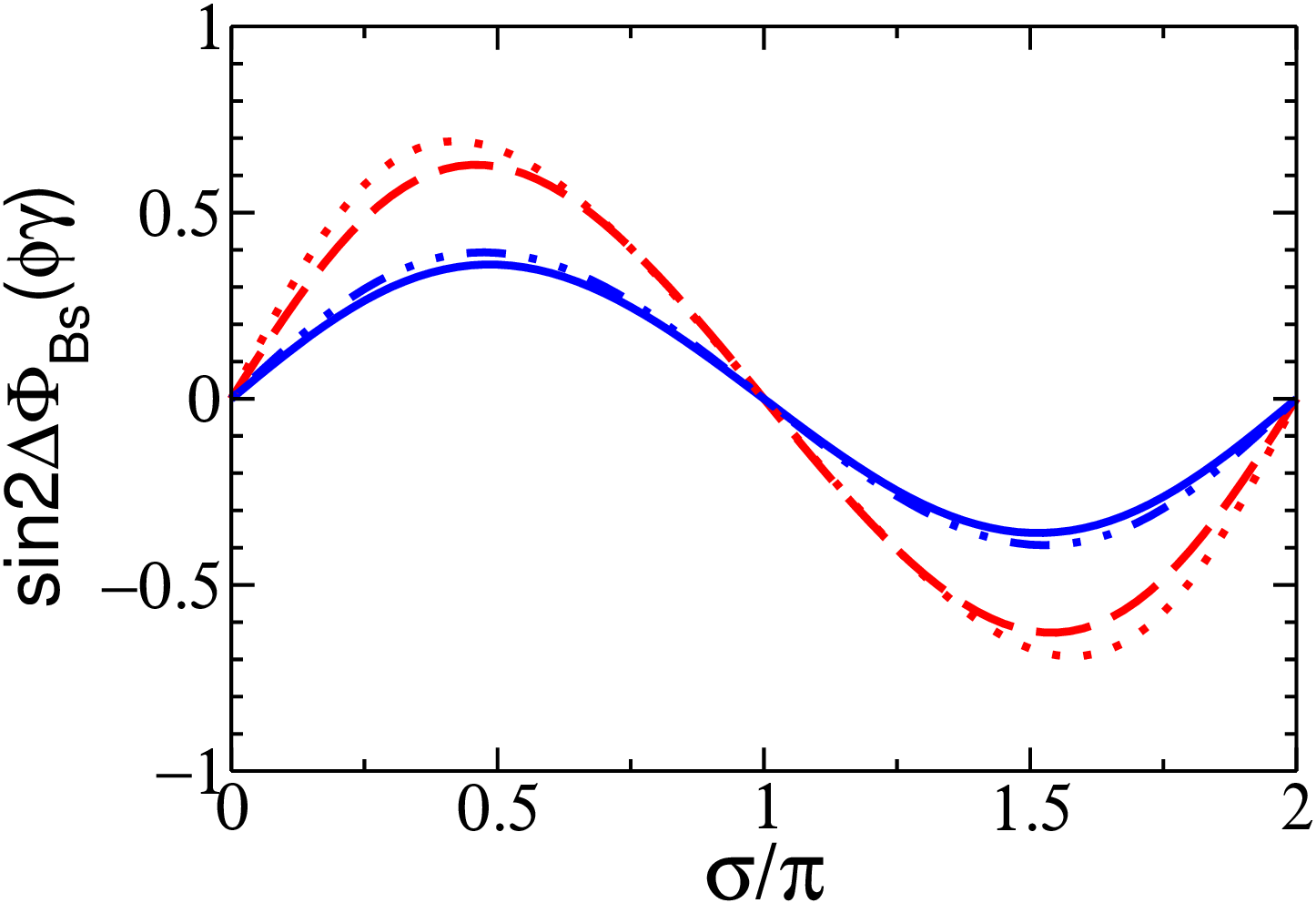}
\hspace{3mm}
\includegraphics[width=1.5in,height=1.3in,angle=0]{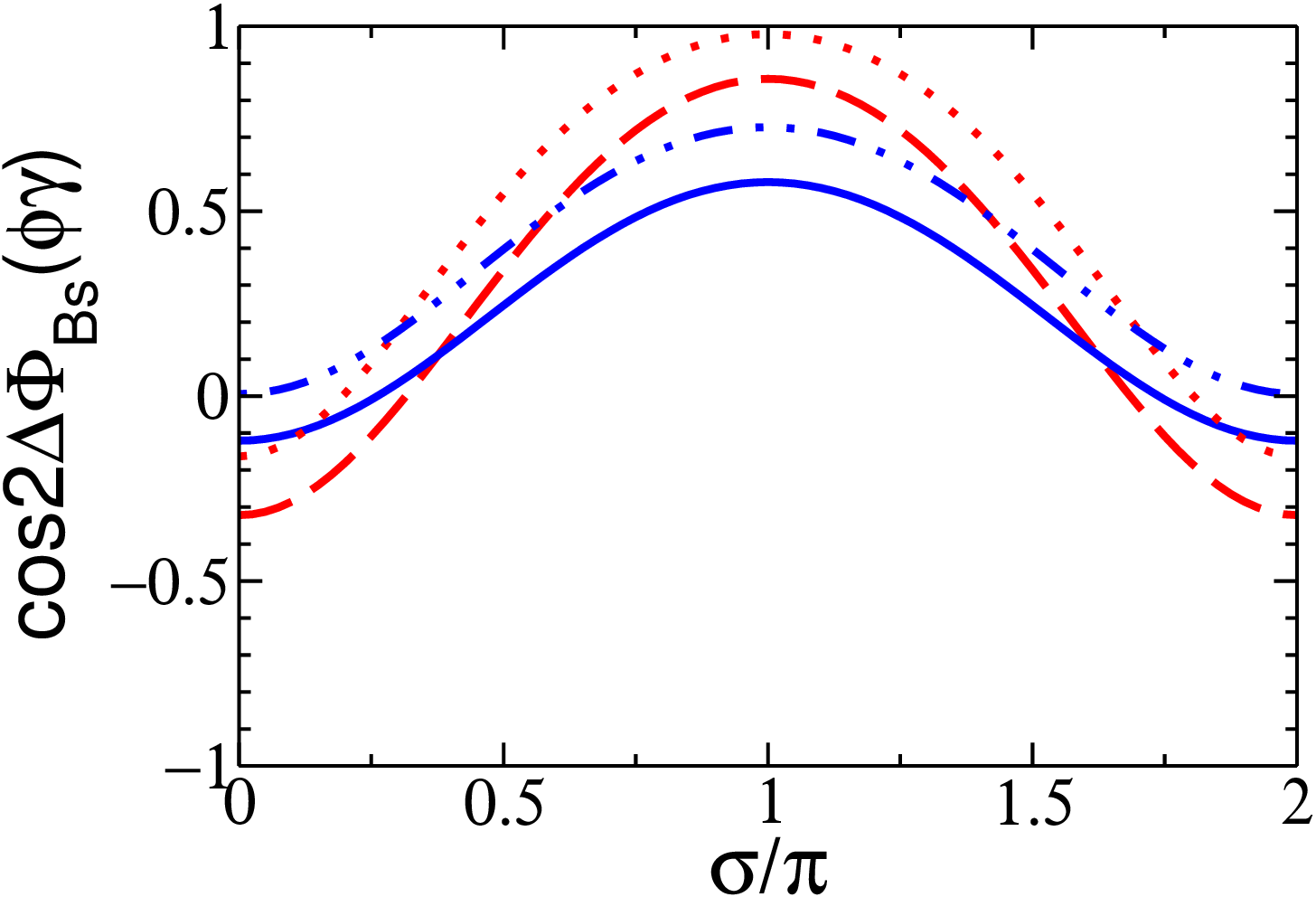}

\vskip 3mm \caption{{\label{fig:phigm}} For $B_s\to \phi\gamma$,
(a)~branching fraction,
(b)~$\sin 2\Phi_{B_s}^{\rm eff}$,
(c)~$\cos 2\Phi_{B_s}^{\rm eff}$,
(d)~$\sin 2\Delta\Phi_{B_s}$ and
(e)~$\cos 2\Delta\Phi_{B_s}$ vs $\sigma$,
with notation same as Fig.\ref{fig:Psiphi}(a).}
\end{figure}

The decay $B_s\to\phi\gamma$ is expected to be the $B_s$
counterpart of the decay $B_d\to K^{\ast 0}\gamma$. The present
experimental upper bound on the rate is \cite{PDG}
\begin{eqnarray}
{\cal B}(B_s\to \phi\gamma) < 1.2\times 10^{-4},
\end{eqnarray}
at $90\%$ confidence level.
In Fig.~\ref{fig:phigm}(a) we illustrate the branching fraction by using
${\cal B}(B_s\to\phi\gamma)=
{\cal B}_{\rm SM}\times
(|c_{11}|^2+|c_{11}^\prime|^2)/|c_{11}^{\rm SM}|^2$,
where $c_{11}^{\rm SM}=-0.31$ and
${\cal B}_{\rm SM} \approx 4.8 \times 10^{-5}$ \cite{Grinstein}.

The quantities of interest are ${\cal S}_{\phi\gamma}$ and ${\cal
A}_{\Delta\Gamma}$. As we will show below, nonvanishing values
imply the presence of {\it wrong helicity photons}. It is known
that ``wrong helicity'' photons from the decay $b\to s\gamma$
would indicate new physics~\cite{AGS,CHH}. In our previous
work~\cite{CHNphik} we have studied ${\cal S}_{K^{\ast
0}(K_S\pi^0)\gamma}$, which was inspired by the recent
experimental data on ${\cal S}_{K_S\pi^0}$ from ${\rm
B_{A}B_{AR}}$ \cite{Browder}. It is important to stress that
${\cal S}_{\phi\gamma}$ and ${\cal A}_{\Delta\Gamma}$ are free
from hadronic effects such as $\tilde S_{K^+K^-}/q^2$ in $B_s\to
K^+K^-$. Hadronic effects are largely absorbed in the $B_s\to
\phi$ form factor, which cancels out when forming CP ratios.

The photon radiated from $\bar B_s(B_s)$ is dominantly
left(right)-handed polarized in the SM. Therefore, in the SM,
Eq.~(\ref{eq:timeCPI}) can be written separately for $\Gamma (\bar
B_{s}(t)\to \phi\gamma_L)$ and $\Gamma (B_{s}(t)\to \phi\gamma_L)$
to good approximation. However, $\phi\gamma_{L(R)}$ is {\it not} a
CP eigenstate {\it if} the photon polarization is measured.
Following Ref.~\cite{AGS}, the amplitudes are given by
\begin{eqnarray}
& &\bar A(\bar B_s\to\phi\gamma_L)= \bar a \cos\Psi e^{i\phi_L},
\nonumber \\
& &\bar A(\bar B_s\to\phi\gamma_R)= \bar a \sin\Psi e^{i\phi_R},
\nonumber \\
& &A(B_s\to\phi\gamma_R)= - a \cos\Psi e^{-i\phi_L},
\nonumber \\
& &A(B_s\to\phi\gamma_L)= - a \sin\Psi e^{-i\phi_R},
\end{eqnarray}
where $\cos\Psi(\sin\Psi)$ is the relative amount of
left(right)-polarized photons, and $\phi_{L(R)}$ is the associated
CP violating phase.
The measurement of time-dependent CP asymmetry for this mode
treats the photon handedness as unmeasured.
Thus, Eq.~(\ref{eq:timeCPI}) should be given by~\cite{AGS,CHH}
\begin{eqnarray}
a_{\rm CP}(t)= \frac{\Gamma (\bar B_{s}(t)\to \phi\gamma) -\Gamma
(B_{s}(t)\to \phi\gamma)} {\Gamma (\bar B_{s}(t)\to \phi\gamma)
+\Gamma (B_{s}(t)\to \phi\gamma)}, \label{eq:timeCPradi}
\end{eqnarray}
where $\Gamma (\bar B_{s}(B_s)\to \phi\gamma)$ sums over the two
separate rates for the final states $\phi\gamma_L$ and
$\phi\gamma_R$. It is in this way that interference and CP
violation can occur.

Assuming $a^2=\bar a^2$, Eq.~(\ref{eq:timeCPradi}) gives
\begin{eqnarray}
{\cal S}_{\phi\gamma}
& = &
\frac{-2|c_{11}c^\prime_{11}|}{|c_{11}|^2+|c^\prime_{11}|^2}
\sin (2\Phi_{B_s}+\phi_L+\phi_R), \nonumber \\
{\cal A}_{\Delta\Gamma}
& = &
\frac{-2|c_{11}c^\prime_{11}|}{|c_{11}|^2+|c^\prime_{11}|^2} \cos
(2\Phi_{B_s}+\phi_L+\phi_R), \label{eq:acp_phigm}
\end{eqnarray}
and ${\cal A}_{\phi\gamma}= 0$. The observables ${\cal
S}_{\phi\gamma}$ and ${\cal A}_{\Delta\Gamma}$ are good probes of
right-handed dynamics, as they vanish with $c_{11}^\prime$.
Note that ${\cal A}_{\phi\gamma}$, ${\cal S}_{\phi\gamma}$ and
${\cal A}_{\Delta\Gamma}$ do not satisfy $|{\cal A}_f|^2 + |{\cal
S}_f|^2 + |{\cal A}_{\Delta\Gamma}|^2 = 1$, because in
Eqs.~(\ref{eq:timeCPradi}) and (\ref{eq:acp_phigm}) one sums over
two distinct components.
In fact,
\begin{equation}
\sqrt{{\cal S}_{\phi\gamma}^2 + {\cal A}_{\Delta\Gamma}^2}
=\frac{2|c_{11}c^\prime_{11}|}{|c_{11}|^2+|c^\prime_{11}|^2}
\equiv \sin2\vartheta_{\rm mix},
\end{equation}
is nothing but the relative interference strength, called
$\sin2\vartheta_{\rm mix}$~\cite{CHH}, between $\bar B_{s}(t)\to
\phi\gamma_L$ and $ \phi\gamma_R$ decay amplitudes induced by
mixing.

Since weak interaction is left-handed, the right-handed effect
from the SM is always accompanied with mass suppression factor.
Hence, ${\cal S}^{\rm SM}_{\phi\gamma} \approx {\cal A}^{\rm
SM}_{\Delta\Gamma} \approx 0$, and $\sin\Delta\Phi_{B_s}$ and
$\cos\Delta\Phi_{B_s}$
also vanish within SM.
We illustrate these quantities in Figs.~\ref{fig:phigm}(b)-(e)
within our scenario. Our results for $\sigma\sim 65^\circ$ are
clearly rather different from the {\it zero} value predicted by
SM. They can be profitably studied, again if  fast $B_S$ mixing
can be overcome. Note that $\cos\Delta\Phi_{B_s}$, which can in
principle be measured without tagging and vertexing, is not far
from zero for $\sigma\sim 65^\circ$ hence not a good discriminant.

\section{\label{sec:discussion}Conclusion}

A hint for new physics has emerged in mixing dependent CP
violation asymmetry in $\bar B_d\to \phi K_S$ decay, which seem to
be of opposite sign to $\bar B_d\to J/\psi K_S$. It is important
to cross check in $B_s$ system.

In an explicit model that combines SUSY and Abelian flavor
symmetry, one has maximal $\tilde s_R$-$\tilde b_R$ squark mixing,
with one new associated CP violating phase $\sigma$. The maximal
mixing could generate one light and flavor mixed
$\widetilde{sb}_1$ squark, which could impact on $b\leftrightarrow
s$ transitions.
The current experimental results, ${\cal S}_{\phi K_S}<0$ while
${\cal S}_{\eta^\prime K_S},\; {\cal S}_{K_S \pi^0}>0$, seem to
suggest $\sigma\sim 65^\circ$ with $m_{\widetilde{sb}_1}=0.2$ TeV
and $m_{\tilde g}\sim 0.5$ TeV. Studies of mixing-dependent CP
asymmetry ${\cal S}_f$ in $B_s$ system in as many modes $f$, is of
great interest to test the model. However, it would be challenging
because of very rapid $B_s$ oscillations implied by the sizable
new physics effect in ${\cal S}_{\phi K_S}<0$. An alternative
approach is to use untagged data to complement the studies of CP
violation in $B_s$ system. This is feasible if
$\Delta\Gamma_{s}/\Gamma_s$ is as large as $O(10\%)$, allowing one
to study CP violating asymmetry in rate, ${\cal
A}_{\Delta\Gamma}$. Thus, it is worthy to investigate CP
asymmetries ${\cal S}_f$ and ${\cal A}_{\Delta\Gamma}$ in the
$B_s$ system.

We have illustrated such a study for $B_s$ system that could
search for new physics effects and assist the determination of
model parameters.
We gave the results for $\bar B_s\to J/\psi\phi$, $K^+K^-$ and
$\phi\gamma$. We stress that ${\cal S}_f$, ${\cal
A}_{\Delta\Gamma}$ and $\Delta\Phi_{B_s}$ (shift in $\Phi_{B_s}$
in specific CP eigenmode relative to $J/\psi\phi$) have good
potential to help pin down the model parameters. However, the
measurement of these quantities may sometimes be challenging
because of fast $B_s$ oscillations. We also emphasize that, the
mechanism that utilizes {\it wrong helicity photon} in $B_s\to
\phi\gamma$ decay allows one to study CP violation without
hadronic effects that plague the charmless hadronic modes such as
$\bar B_s\to K^+K^-$.

\begin{acknowledgments}
M.N. would like to thank F. Ukegawa for useful discussions.
This work is supported in part by grants
NSC-92-2811-M-001-054 and NSC92-2811-M-002-033, the BCP Topical
Program of NCTS, and the MOE CosPA project.
\end{acknowledgments}

\end{document}